\documentclass[journal]{IEEEtran}
\IEEEoverridecommandlockouts
\usepackage{cite}
\usepackage{amsmath,amssymb,amsfonts}
\usepackage{graphicx}
\usepackage{textcomp}
\usepackage{xcolor}
\usepackage{algorithm}
\usepackage{algorithmic}
\usepackage{enumitem}
\usepackage{footnote}
\usepackage{multirow}
\usepackage{threeparttable}
\usepackage{makecell}
\usepackage{booktabs}
\usepackage[colorlinks,linkcolor=blue]{hyperref}
\usepackage{subcaption}

\usepackage{array}

\def\BibTeX{{\rm B\kern-.05em{\sc i\kern-.025em b}\kern-.08em
    T\kern-.1667em\lower.7ex\hbox{E}\kern-.125emX}}

\makeatletter
\renewcommand{\maketag@@@}[1]{\hbox{\m@th\normalsize\normalfont#1}}%
\makeatother

\begin{document}

\title{Lossy Microwave Linear Analog Computer (MiLAC) for Future MIMO: \\Learning-based Architecture Designs for Spectral and Energy Efficiency Maximization}
\author{Binggui Zhou,~\IEEEmembership{Member,~IEEE},
        and Bruno Clerckx,~\IEEEmembership{Fellow,~IEEE}
\thanks{Binggui Zhou and Bruno Clerckx are with the Department of Electrical and Electronic Engineering, Imperial College London, London SW7 2AZ, United Kingdom. (e-mails:\{binggui.zhou,~b.clerckx\}@imperial.ac.uk).}
}

\maketitle

\begin{abstract}
Microwave linear analog computers (MiLACs) offer a transformative paradigm for future multiple-input multiple-output (MIMO) systems by shifting complex signal processing into the analog domain, thereby significantly reducing computational complexity, radio-frequency chains, and analog-digital converters, while speeding up computation. However, the practical deployment of MiLACs is severely constrained by the inherent hardware losses of the tunable admittance components (TACs) interconnecting MiLAC ports, which introduce severe inter-stream interference and fundamentally limit the spectral efficiency (SE) of the system. In addition, while denser architectures offer greater spatial degrees of freedom to mitigate inter-stream interference, the cumulative hardware losses and power consumption of massive TACs severely degrade the system's energy efficiency (EE). Consequently, designing architectures for lossy MiLACs emerges as a critical yet unresolved challenge, as it necessitates striking a delicate tradeoff between interference suppression and cumulative hardware losses/power consumption. To address this challenge, this paper investigates the joint MiLAC architecture design and performance (SE/EE) maximization in lossy MiLAC-aided MIMO systems. We propose a novel learning-based joint architecture and performance optimization framework (LJAPOF) that unifies the design of MiLAC architectures and analog beamforming configurations for lossy MiLACs under both SE- and EE-oriented objectives. Numerical results demonstrate that by intelligently navigating the fundamental tradeoff between interference suppression and hardware/power consumption, the proposed LJAPOF can design optimal MiLAC architectures that consistently outperform stem-connected and fully-connected MiLACs in maximizing the system's SE and EE.

\end{abstract}


\section{Introduction}

Multiple-input multiple-output (MIMO) has been a key enabling technology in fifth-generation (5G) networks and is expected to remain fundamental in sixth-generation (6G) systems, providing rich spatial degrees of freedom (DoF) and substantial multiplexing gains. However, realizing the full potential of increasingly large MIMO presents a severe challenge: the high power consumption and hardware cost of conventional fully-digital transceivers arising from radio frequency (RF) chains constructed by high-resolution analog-to-digital converters (ADCs) / digital-to-analog converters (DACs) and mixers. Consequently, overcoming this challenge requires a fundamental paradigm shift from digital-domain signal processing to analog-domain signal processing\cite{caloz2013analog}. By executing matrix operations and beamforming directly in the analog domain, analog computing significantly reduces the reliance on costly and energy-hungry digital components, which enables scalable and energy-efficient MIMO deployments. Recently, microwave linear analog computers (MiLACs) have been proposed as a practical approach to realize signal processing directly in the analog domain using multiport microwave networks with tunable admittance components (TACs)\cite{nerini2025analog1}. Leveraging the analog-domain signal processing capabilities, MiLACs can significantly reduce computational complexity and have been applied to enable efficient precoding and combining in gigantic MIMO systems\cite{nerini2025analog2,nerini2025capacity,wu2026microwave,nerini2026physics,zhang2026channel}.

Regardless of the immense theoretical promise of MiLACs, existing literature relies on the optimistic assumption of lossless TACs \cite{nerini2025capacity, nerini2026mimo}, limiting the practical deployment of MiLACs. Prior studies on TACs and reconfigurable microwave networks, particularly in the context of beyond-diagonal reconfigurable intelligent surfaces (BD-RISs) \cite{shen2022modeling,li2024reconfigurable,li2026tutorial, wu2025beyonddiagonal, wu2026beyonddiagonal}, have demonstrated that the inherent hardware losses of TACs impose fundamental limitations on system performance \cite{zhou2025beyonddiagonal,peng2026lossy}. Unlike lossy BD-RISs, which merely manipulate the propagation environment and still rely on the fully digital basebands of the transceivers to orthogonalize data streams and execute power allocation, lossy MiLACs function directly as the analog transceiver front-ends. As a consequence, the hardware losses in lossy MiLACs physically induce severe inter-stream interference, fundamentally capping the spatial multiplexing gain and ultimately restricting the system’s spectral efficiency (SE).

In addition, previous works on MiLAC-aided beamforming primarily rely on fully-connected MiLAC architectures, where all ports are interconnected via TACs to maximize spatial DoF for precise beamforming and robust interference suppression \cite{nerini2025analog2,nerini2025capacity,wu2026microwave}. However, the circuit complexity (i.e., the number of TACs / interconnections among MiLAC ports) of fully-connected MiLACs scales quadratically with both the number of antennas and the spatial multiplexing order, leading to prohibitive power consumption and poor energy efficiency (EE) in large-scale deployments. Recent studies have introduced stem-connected MiLACs \cite{nerini2026mimo} to drastically reduce the circuit complexity without compromising the system's SE, though they remain fundamentally limited by their validation solely on ideal MiLACs. Furthermore, although an algorithm to maximize the EE of MiLAC-aided systems was proposed in \cite{zhang2026quantization}, it strictly assumes a lossless fully-connected architecture where both circuit complexity and power consumption are rigidly fixed.

Despite existing efforts to either account for the losses of TACs in BD-RISs or mitigate the excessive circuit complexity of ideal MiLACs, actively designing the architectures of lossy MiLACs to strike a delicate tradeoff between interference suppression and cumulative hardware loss/power consumption emerges as a critical yet unresolved challenge. While increasing interconnections theoretically offers enhanced spatial DoF to suppress inter-stream interference and sustain higher multiplexing gains, it simultaneously introduces escalating hardware losses and power consumption. Consequently, the net impact of architectural densification is highly ambiguous, as the anticipated multiplexing gains can be easily overshadowed by these cumulative losses, degrading the system's spectral and energy efficiency.

Resolving this ambiguity necessitates the joint optimization of the discrete MiLAC architecture (i.e., pruning or activating interconnections) and the continuous analog beamforming configurations. However, this inherently formulates a highly intractable mixed-integer non-linear programming (MINLP) problem\cite{sahinidis2019mixedinteger}. Conventional optimization techniques often struggle with the combinatorial explosion of the architectural search space and the severe non-convexity introduced by TAC hardware constraints, leading to prohibitive computational overhead and locally suboptimal solutions. Recently, machine learning (ML) has emerged as a disruptive paradigm for physical layer design in MIMO systems \cite{zhou2024lowoverhead,peng2026joint,xiao2026deep}. By leveraging the powerful approximation capabilities of neural networks, learning-based approaches can efficiently navigate vast, highly non-convex solution spaces and directly map channel states to optimal analog beamforming configurations in real-time, offering a compelling mechanism to overcome the bottlenecks of conventional iterative algorithms \cite{zhou2025beyond}.

Motivated by these unresolved challenges, this paper systematically investigates the joint architecture design and performance (SE/EE) optimization in lossy MiLAC-aided MIMO systems by pioneering a novel learning-based framework. Consequently, our study establishes rigorous design guidelines that prevent the deployment of blindly over-engineered MiLAC architectures, which otherwise suffer from severe capacity collapse and profound energy inefficiency under real-world physical constraints. The major contributions of this paper are summarized as follows:

\begin{enumerate}

\item We formulate two joint architecture design and performance optimization problems to design optimal lossy MiLAC architectures that maximize the system's SE and EE, respectively, as the optimal tradeoff between interference suppression and cumulative hardware losses/power consumption shifts significantly depending on the performance objective. To solve these problems, we propose a novel learning-based joint architecture and performance optimization framework (LJAPOF). Serving as a universal solution for both SE- and EE-oriented lossy MiLAC architecture designs, the LJAPOF explicitly incorporates the physical loss models of TACs to jointly optimize the architecture and analog beamforming configurations in an end-to-end manner, thereby elegantly transforming a highly intractable non-convex optimization problem into a resolvable unsupervised learning task.

\item The proposed LJAPOF comprises a differentiable MiLAC architecture generator for architecture design, a physics-informed capacitance learning network (PICLN) for physics-informed analog beamforming, a physics-constrained differentiable water-filling (PCDWF) layer for optimal power allocation, and a dual-rate loss structure for end-to-end unsupervised learning. The differentiable MiLAC architecture generator enables intelligent pruning of MiLAC interconnections to dynamically tailor the architecture driven directly by the targeted objective. The PICLN effectively bridges the gap between data-driven deep learning and rigorous microwave network theory, utilizing a residual-based architecture and SVD-guided features to stably learn real-time, hardware-constrained capacitance values for optimal analog beamforming. The PCDWF layer enables interpretable, physics-grounded, and parameter-free optimal power allocation by natively integrating the water-filling algorithm into the differentiable computational graph, showing parameter and training complexity advantages over existing deep learning-based power allocation methods. The dual-rate loss structure incorporates a shadow achievable rate, derived via conventional water-filling, to serve as a continuous gradient-preserving anchor that prevents severe shortcut learning and the single-stream trap, ultimately ensuring robust spatial multiplexing.

\item Numerical results demonstrate the effectiveness of the proposed LJAPOF and highlight several interesting observations. \textit{First}, while the SE of ideal MiLAC-aided systems grows and eventually saturates with an increasing number of streams $N_S$, that of lossy MiLAC-aided systems experiences an initial improvement before ultimately declining. This reversal occurs because the diminishing marginal multiplexing gains are eventually overshadowed by the escalating hardware losses induced by increased interconnections. Consequently, stem-connected MiLACs inherently achieve higher SE than fully-connected architectures in lossy environments due to their reduced hardware losses. \textit{Second}, unlike the settings in previous works, e.g., \cite{nerini2025analog2, nerini2025capacity, nerini2026mimo}, where the transmitter-side and receiver-side MiLACs have the same stem-connected/fully-connected architectures, the results reveal that intelligently synthesizing objective-driven, asymmetrically sparse transmitter-side and receiver-side MiLAC architectures is paramount to overcoming the severe capacity and power bottlenecks imposed by physical hardware losses, regardless of whether these MiLAC architectures are obtained via SE or EE optimization. Specifically, via SE optimization, the transmitter-side MiLAC becomes sparse-connected, while the receiver-side MiLAC remains fully-connected. Consequently, different from stem-connected and fully-connected lossy MiLAC-aided systems, the SE of the SE-oriented MiLAC-aided system first increases and then stabilizes as $N_S$ scales up. While via EE optimization, both the transmitter-side and receiver-side MiLAC become sparse-interconnected, the latter one always maintains a denser architecture than the former one. \textit{Third}, when $N_S$ is high (e.g., $N_S=12$ and $N_S=16$ when the number of transmitter and receiver antennas $N_T=N_R=32$), both the SE-oriented and EE-oriented designs successfully synthesize high-performance architectures that substantially outperform the baselines. Specifically, compared with stem-connected MiLACs, the SE-oriented MiLACs achieve $4.4\%$ and $9.0\%$ gains in SE, along with $15.7\%$ and $42.1\%$ improvements in EE, respectively. In contrast, the EE-oriented MiLACs deliver substantial EE gains of $80.1\%$ and $162.4\%$, respectively, along with comparable or slightly higher SE. Beyond demonstrating the effectiveness of the proposed joint designs, these results explicitly illuminate the fundamental EE-SE tradeoff inherent in lossy MiLAC-aided MIMO systems. By directly comparing the two optimization orientations, a compelling tradeoff emerges: for $N_S \in \{4,8,12,16\}$, the EE-oriented MiLACs significantly boost the EE by $29.1\%$ to $84.7\%$ over the SE-oriented counterparts, at the cost of a mere $2.5\%$ to $6.6\%$ sacrifice in SE.

\end{enumerate}

\textit{Organization}: In Section \ref{Sec. Sys.}, we introduce the lossy MiLAC-aided MIMO system and formulate the joint architecture and SE/EE optimization problems. In Section \ref{Sec. Method}, we propose the LJAPOF. Section \ref{Sec. Sim.} presents simulation results to demonstrate the effectiveness of the proposed LJAPOF and reveal optimal lossy MiLAC architectures that maximize the SE and EE, respectively. Finally, Section \ref{Sec. Con.} concludes this work.

\textit{Notation}: Bold uppercase letter $\mathbf{A}$ and bold lowercase letter $\mathbf{a}$ represent a matrix and a vector, respectively. Calligraphy uppercase letter $\mathcal{A}$ represents a set. $[\mathbf{A}]_{m,:}$ and $[\mathbf{A}]_{m,n}$ denote the $m$-th row and the element at the $m$-th row and $n$-th column of $\mathbf{A}$, respectively. $(\cdot)^T$, $(\cdot)^H$, and $(\cdot)^{-1}$ denote the transpose, conjugate-transpose, and inverse of a matrix, respectively. $\mathbb{E}[\cdot]$ denotes the expectation operator. $\Vert\cdot\Vert_0$, $\Vert\cdot\Vert_2$, and $\Vert\cdot\Vert_F$ denote the L0 norm, L2 norm, and Frobenius norm, respectively. $\Re\{\cdot\}$ and $\Im\{\cdot\}$ take the real and imaginary parts of the input, respectively. $\odot$ denotes the Hadamard product. $\mathbb{I}(\cdot)$ represents the indicator function, and $\operatorname{diag}(\mathbf{a})$ denotes a diagonal matrix with the elements of vector $\mathbf{a}$ on its main diagonal.

\section{Lossy MiLAC-Aided MIMO System}\label{Sec. Sys.}

\subsection{System Model}
A MiLAC is a $P$-port reconfigurable microwave network made of TACs interconnecting its ports and capable of realizing signal processing directly in the analog domain\cite{nerini2025analog1}. The MiLAC can be characterized through its admittance matrix $\mathbf{Y} \in \mathbb{C}^{P \times P}$ (also known as the Y-parameter) according to multiport network theory\cite{pozar2011microwave}. Based on the individual admittance values of TACs $\{Y_{i,j}\}_{i,j=1}^{P}$, the entries of $\mathbf{Y}$ are given by
\begin{equation}
\mathbf{Y}_{i,j}=
\begin{cases}
-Y_{i,j} & i\neq j\\
\sum_{p=1}^PY_{p,j} & i=j
\end{cases},\label{eq:Yik-entry}
\end{equation}
for $i,j=1,\ldots,P$.
Considering a reciprocal reconfigurable microwave network, we have $\mathbf{Y}=\mathbf{Y}^T$.

\begin{figure*}[htbp]
\centering
\includegraphics[width=0.8\textwidth]{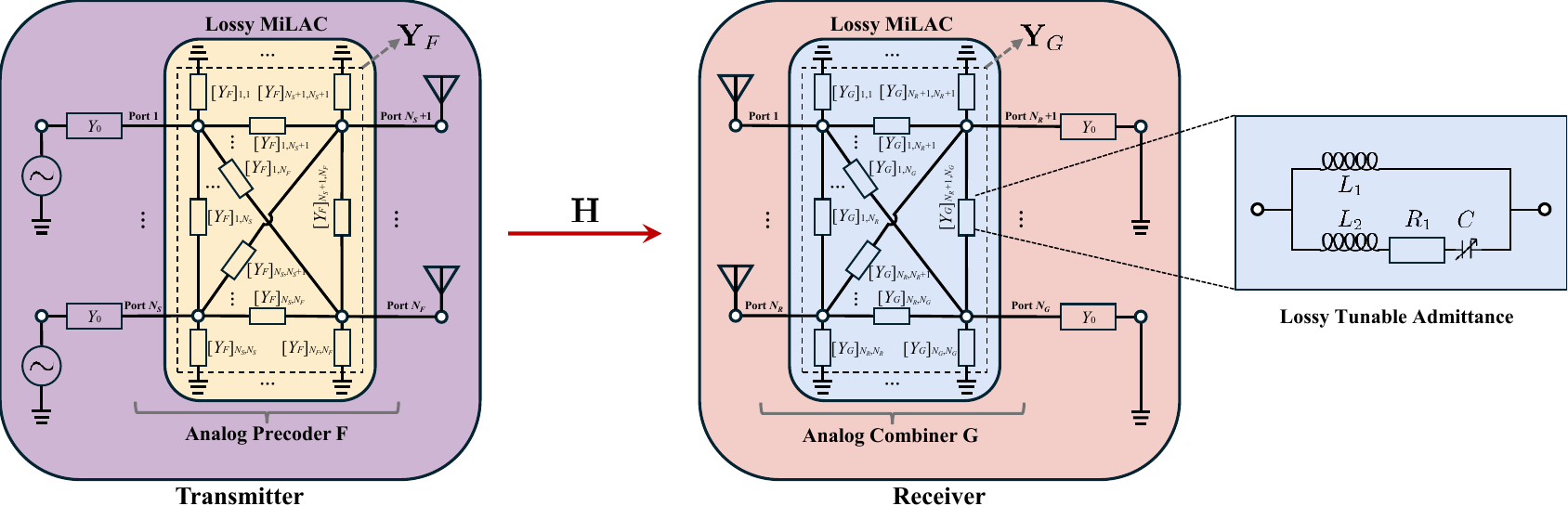}
\caption{The lossy MiLAC-aided point-to-point MIMO system.}
\label{system}
\end{figure*}

In this paper, we consider a point-to-point MIMO system aided by a MiLAC-based precoder at the transmitter side and a MiLAC-based combiner at the receiver side, respectively, as depicted in Fig. \ref{system}. The transmitter is equipped with $N_T$ antennas, and the receiver is equipped with $N_R$ antennas. The transmitter transmits $N_S$ independent data streams to the receiver, with $N_S \leq \min\{N_T, N_R\}$. The transmitter-side and receiver-side MiLACs have $N_F = N_S + N_T$ and $N_G = N_R + N_S$ ports, respectively. Let $P_T$ denote the total transmit power and $\mathbf{p} = [p_1, \ldots, p_{N_S}]^T \in \mathbb{R}^{N_S \times 1}$ denote the power allocation vector, where $p_s \ge 0$ represents the power allocation factor assigned to the $s$-th data stream, subject to $\sum_{s=1}^{N_S} p_s = 1$. In this system, signal processing is natively executed in the analog domain by the respective MiLACs. Specifically, the transmitter-side MiLAC synthesizes the precoding matrix $\mathbf{F} \in \mathbb{C}^{N_T \times N_S}$, while the receiver-side MiLAC applies the combining matrix $\mathbf{G} \in \mathbb{C}^{N_S \times N_R}$. Denoting $\mathbf{s} = [s_1, \ldots, s_{N_S}]^T \in \mathbb{C}^{N_S \times 1}$ as the normalized data symbol vector satisfying $\mathbb{E}[\mathbf{s}\mathbf{s}^H] = \mathbf{I}_{N_S}$, the signal vector $\mathbf{z} \in \mathbb{C}^{N_S \times 1}$ after the receiver-side MiLAC processing can be formulated as
\begin{equation}
\mathbf{z} = \underbrace{\sqrt{P_T} \mathbf{G} \mathbf{H} \mathbf{F} \operatorname{diag}(\sqrt{\mathbf{p}}) \mathbf{s}}_{\bar{\mathbf{z}}} + \underbrace{\mathbf{G} \mathbf{n}}_{\bar{\mathbf{n}}}, \label{signal}
\end{equation}
where $\mathbf{H} \in \mathbb{C}^{N_R \times N_T}$ denotes the wireless channel and $\mathbf{n} \in \mathbb{C}^{N_R \times 1}$ accounts for the receive antenna noise vector characterized by the covariance matrix $\mathbb{E}[\mathbf{n}\mathbf{n}^H] = \sigma^2 \mathbf{I}_{N_R}$, with $\sigma^2$ representing the noise power, and $\bar{\mathbf{z}}$ and $\bar{\mathbf{n}}$ denote the noise-free signal vector and the combined noise, respectively.

According to \cite{nerini2025analog2}, the precoding matrix $\mathbf{F}$ is a function of the admittance matrix of the transmitter-side MiLAC $\mathbf{Y}_F\in\mathbb{C}^{N_F \times N_F}$ as
\begin{equation}
\mathbf{F}=\left[\left(\frac{\mathbf{Y}_F}{Y_0}+\mathbf{I}_{N_F}\right)^{-1}\right]_{N_S+1:N_S+N_T, 1:N_S},\label{F}
\end{equation}
where $Y_0$ denotes the reference admittance. Similarly, the combining matrix $\mathbf{G}$ is a function of the admittance matrix of the receiver-side MiLAC $\mathbf{Y}_G\in\mathbb{C}^{N_G \times N_G}$ as
\begin{equation}
\mathbf{G}=\left[\left(\frac{\mathbf{Y}_G}{Y_0}+\mathbf{I}_{N_G}\right)^{-1}\right]_{N_R+1:N_R+N_S, 1:N_R}.\label{G}
\end{equation}

\subsection{Modeling of Lossy MiLACs}
To accurately capture the hardware losses of TACs in lossy MiLACs, we adopt the lossy TAC model in \cite{peng2026lossy}. Specifically, as shown in Fig. \ref{system}, a TAC has two inductances with fixed values, $L_1$ and $L_2$, a parasitic resistance $R_1$ with a fixed value that accounts for power dissipation, and a tunable capacitance $C$. Consequently, in contrast to a lossless MiLAC with purely susceptive TACs\cite{nerini2025analog2,nerini2025capacity,wu2026microwave,nerini2026physics,zhang2026channel}, the TAC of a lossy MiLAC necessitates a complex representation, $Y = G + jB$, accounting for both its real part, i.e., the conductance $G$, and imaginary part, i.e., the susceptance $B$. Based on this, the individual TACs of the transmitter-side and receiver-side MiLACs, i.e., $\{[Y_F]_{i,j} = [G_F]_{i,j} + j[B_F]_{i,j}\}_{i,j=1}^{N_F}$ and $\{[Y_G]_{i,j} = [G_G]_{i,j} + [B_G]_{i,j}\}_{i,j=1}^{N_G}$, can be modeled with
\begin{subequations}
\begin{align}
[G_F]_{i,j} &= \frac{R_1}{R_1^2 + \left(\omega L_2 - \frac{1}{\omega [C_F]_{i,j}} \right)^2}, \label{R_YF}\\
[B_F]_{i,j} &= -\frac{1}{\omega L_1} + \frac{-\omega L_2 + \frac{1}{\omega [C_F]_{i,j}}}{R_1^2 + \left(\omega L_2 - \frac{1}{\omega [C_F]_{i,j}} \right)^2}, \label{I_YF}
\end{align}
\end{subequations}
and
\begin{subequations}
\begin{align}
[G_G]_{i,j} &= \frac{R_1}{R_1^2 + \left(\omega L_2 - \frac{1}{\omega [C_G]_{i,j}} \right)^2}, \label{R_YG}\\
[B_G]_{i,j} &= -\frac{1}{\omega L_1} + \frac{-\omega L_2 + \frac{1}{\omega [C_G]_{i,j}}}{R_1^2 + \left(\omega L_2 - \frac{1}{\omega [C_G]_{i,j}} \right)^2}, \label{I_YG}
\end{align}
\end{subequations}
respectively, where $\{[C_F]_{i,j}\}_{i,j=1}^{N_F}$ and $\{[C_G]_{i,j}\}_{i,j=1}^{N_G}$ denote the corresponding capacitance values of TACs constituting the transmitter-side and receiver-side MiLACs, respectively, and $\omega = 2\pi f$ denotes the angular frequency with $f$ being the signal frequency.

\subsection{Architecture and Circuit Complexity of a Lossy MiLAC}
To mathematically characterize the architecture and circuit complexity of a MiLAC, we define a binary architecture characterization matrix. The diagonal entries of this matrix capture the connections from individual MiLAC ports to ground via TACs, while the off-diagonal entries capture the interconnections between distinct MiLAC ports via TACs. Specifically, an entry of $1$ indicates the presence of a TAC connection, whereas $0$ indicates no connection. The circuit complexity of the MiLAC can then be specified as the total number of nonzero entries in the lower-triangular part of the architecture characterization matrix.

Denoting $\mathbf{A}_F$ and $\mathbf{A}_G$ as the architecture characterization matrices of the transmitter-side and receiver-side MiLACs, respectively, $\mathbf{Y}_F$ and  $\mathbf{Y}_G$ can then be characterized by
\begin{align}
\label{matY_F}
& \left[\mathbf{Y}_F\right]_{i,j} = \begin{cases}
-[Y_F]_{i,j}, &\left[\mathbf{A}_F\right]_{i,j} = 1 \text{ and } i\ne j , \\
\sum_{k=1}^{N_F}[Y_F]_{i,k}, &\left[\mathbf{A}_F\right]_{i,j} = 1 \text{ and }  i = j, \\
0,  & \left[\mathbf{A}_F\right]_{i,j} = 0,
\end{cases}
\end{align}
and
\begin{align}
\label{matY_G}
& \left[\mathbf{Y}_G\right]_{i,j} = \begin{cases}
-[Y_G]_{i,j}, &\left[\mathbf{A}_G\right]_{i,j} = 1 \text{ and } i\ne j , \\
\sum_{k=1}^{N_G}[Y_G]_{i,k}, &\left[\mathbf{A}_G\right]_{i,j} = 1 \text{ and }  i = j, \\
0,  & \left[\mathbf{A}_G\right]_{i,j} = 0,
\end{cases}
\end{align}
respectively. The circuit complexity of the transmitter-side MiLAC and the receiver-side MiLAC can be specified as 
\begin{align}
K_{F} &= \sum_{i=1}^{N_F} \sum_{j=1}^{i} \left[\mathbf{A}_F\right]_{i,j}, \label{K_F}\\
K_{G} &= \sum_{i=1}^{N_G} \sum_{j=1}^{i} \left[\mathbf{A}_G\right]_{i,j} \label{K_G}.
\end{align}

\subsection{Problem Formulation}

To enable architecture design for lossy MiLACs that maximizes the SE and EE of lossy MiLAC-aided MIMO systems, we formulate two joint architecture and performance optimization problems, i.e., the SE- and EE-oriented joint architecture and performance optimization problems, respectively.

\subsubsection{SE-oriented Joint Architecture Design and Performance Optimization}
Building upon \eqref{signal} and denoting the effective channel as $\mathbf{E} = \mathbf{G}\mathbf{H}\mathbf{F}\in\mathbb{C}^{N_S \times N_S}$, the SINR for the $s$-th data stream is defined as
\begin{equation}
\text{SINR}_s = \frac{P_T p_s \left\vert [\mathbf{E}]_{s,s} \right\vert^2}{P_T \sum_{t \neq s} p_t \left\vert [\mathbf{E}]_{s,t} \right\vert^2 + \left\Vert [\mathbf{G}]_{s,:} \right\Vert^2 \sigma^2}, \label{sinr}
\end{equation}
where the numerator represents the desired signal power, and the denominator aggregates the inter-stream interference arising from the off-diagonal elements of $\mathbf{E}$ and the effective noise power amplified by the analog combiner.
Then, the system's SE can be evaluated by aggregating the achievable rate of each individual stream, which is formulated as
\begin{equation}
R=\sum_{s=1}^{N_S}\log_2\left(1+\text{SINR}_s\right).\label{rate}
\end{equation}

As proved in \cite{nerini2025capacity}, lossless and reciprocal MiLAC-aided beamforming can achieve the optimal SE (i.e., the capacity)
\begin{equation}
R^{\star}=\sum_{s=1}^{N_S}\log_2\left(1+\frac{P_Tp_{s}^\star\lambda_{s}}{4\sigma^2}\right),\label{capacity}
\end{equation}
where $\lambda_s$ represents the $s$-th eigenvalue of the channel covariance matrix $\mathbf{H}\mathbf{H}^H$ and $p_s^\star$ denotes the power allocation factor for the $s$-th stream obtained through water-filling \cite{clerckx2013mimo} as
\begin{equation}
p_s^\star=\max\left\{0,\mu-\frac{4\sigma^2}{P_T\lambda_s}\right\},\label{water-filling solution}
\end{equation}
where $\mu$ is selected to satisfy $\sum_{s=1}^{N_S}p_{s}=1$.

However, as previously discussed, a lossy MiLAC physically induces severe inter-stream interference, making $R^{\star}$ in \eqref{capacity} no longer achievable in a lossy MiLAC-aided system. Therefore, to maximize the system's SE, the highly non-convex SE-oriented joint architecture and performance optimization problem needs to be formulated considering all architectural (i.e., \eqref{matY_F} and \eqref{K_F} / \eqref{matY_G} and \eqref{K_G}), lossy (i.e., \eqref{R_YF} and \eqref{I_YF} / \eqref{R_YG} and \eqref{I_YG}), and analog computation (i.e., \eqref{F} / \eqref{G}) constraints of lossy MiLACs, which writes as
\begin{subequations}
\begin{align}
(\mathcal{P}_1): \quad & \max_{\mathbf{F}, \mathbf{G}, \mathbf{p}}
& & R \\
& \text{s.t.}
& & \sum_{s=1}^{N_S} p_s = 1, \\
&&& K_{F} \leq K_{F,\max},  ~ K_{G} \leq K_{G,\max}, \\
&&& \mathbf{F} \in \mathcal{F}, ~ \mathbf{G} \in \mathcal{G},
\end{align}
\end{subequations}
where $K_{F,\max}$ and $K_{G,\max}$ denote the maximum acceptable circuit complexity for the transmiter-side and the receiver side MiLACs, respectively, and $\mathcal{F}$/$\mathcal{G}$ denotes the feasible set of the analog precoder/combiner under the architectural, lossy, and analog computing constraints of lossy MiLACs.

It is worth emphasizing that according to \eqref{sinr}, the inherent cumulative hardware losses of lossy TACs in the transmitter-side MiLAC directly attenuate the transmit signal power, thereby severely degrading the SE. As interconnection density increases at the transmitter-side MiLAC, the diminishing marginal SE gains are eventually eclipsed by steadily escalating cumulative hardware losses, rendering the resulting net SE benefit negative. To prevent this profound signal energy leakage from overshadowing the SE gains, the SE optimization inherently leads to sparse transmitter-side MiLAC architectures with low circuit complexity. Conversely, according to \eqref{signal}, since the receiver-side MiLAC performs combining entirely in the analog domain, the receive antenna noise $\mathbf{n}$ is also processed by the combiner $\mathbf{G}$. Therefore, the hardware losses in the receiver-side MiLAC attenuate the noise alongside the signal, leading to a substantially lower impact on the system's SE compared to that at the transmitter side. Consequently, the SE optimization will converge toward MiLAC architectures that are significantly more complex (i.e., with higher circuit complexity) at the receiver side while remaining highly sparse at the transmitter side.\footnote{The asymmetry in the circuit complexity of the transmitter-side and receiver-side MiLACs has also been numerically verified in Section \ref{Sec. Sim. SE_EE}.}

\subsubsection{EE-oriented Joint Architecture Design and Performance Optimization}

Since $(\mathcal{P}_1)$ strictly targets SE maximization, it inherently neglects the power consumption of MiLAC architectures, yielding designs that are prohibitively energy-inefficient. Therefore, to realize a practical MIMO system with highly energy-efficient analog precoding and combining, we need to further formulate an EE-oriented joint architecture and performance optimization problem that explicitly incorporates the power consumption of lossy MiLACs into the objective. Specifically, we define the total power consumption of the system as
\begin{align}
&P_{\text{total}} = \ \eta^{-1}P_T + P_{\text{circ},T}(N_{\text{act},T}) + P_{\text{circ},R}(N_{\text{act},R})  \notag \\
& + \underbrace{P_{\text{dc},F}(K_F) + P_{\text{ohmic},F}(\mathbf{Y}_F)}_{P_{\text{total},F}} + \underbrace{P_{\text{dc},F}(K_G) + P_{\text{ohmic},G}(\mathbf{Y}_G)}_{P_{\text{total},G}},
\end{align}
where $\eta$ denotes the power added efficiency of the transmit power amplifier, $P_{\text{circ},T}(N_{\text{act},T})$ and $P_{\text{circ},R}(N_{\text{act},R})$ account for the power consumption of the circuits (excluding the lossy MiLAC-related circuits) in the transmitter and receiver, respectively, and $P_{\text{total},F}$ and $P_{\text{total},G}$ denote the total power consumption of the transmitter-side and receiver-side lossy MiLAC, respectively. Here, $N_{\text{act},T} = \Vert\mathbf{p}\Vert_0 \le N_S$ denotes the number of transmitter-side active RF chains determined by the power allocation $\mathbf{p}$, and $N_{\text{act},R} = \sum_{m=1}^{N_S} \mathbb{I} \left( [\mathbf{p}_{rx}]_m \ge \xi \cdot \max_{k} [\mathbf{p}_{rx}]_k \right)$ denotes the number of active RF chains at the receiver side, where $\mathbb{I}(\cdot)$ refers to the indicator function, $\mathbf{p}_{rx} = \operatorname{diag}(\mathbf{R}_{\bar{z}})$ refers to the received signal power, and $\mathbf{R}_{\bar{z}} = P_T \mathbf{G} \mathbf{H} \mathbf{F} \operatorname{diag}(\mathbf{p}) \mathbf{F}^H \mathbf{H}^H \mathbf{G}^H$ represents the covariance matrix of $\bar{\mathbf{z}}$. To avoid activating RF chains with negligible received signal power, the receiver-side baseband controller adopts an effective dynamic range thresholding strategy.\footnote{In this paper, we assume that the transceivers can dynamically adapt the number of active RF chains on a per-coherence-block basis, driven by the instantaneous power allocation. It is worth noting that modern RF transceivers can be designed to support microsleep modes\cite{lauridsen2016sleep}, where the biasing currents of power-hungry components (e.g., ADCs/DACs, LNAs, and mixers) can be toggled within a few microseconds ($\mu\text{s}$). Since the channel coherence time is typically on the order of milliseconds ($\text{ms}$), the transient switching latency is practically negligible compared to the data transmission payload duration, rendering instantaneous RF chain activation highly feasible in 5G-Advanced and 6G systems.} Specifically, a receive RF chain is activated only if its captured signal power $[\mathbf{p}_{rx}]_m$ is no less than a fraction $\xi \in (0, 1)$ of the maximum spatial stream power $\max_{k} [\mathbf{p}_{rx}]_k$. $P_{\text{circ},T}(N_{\text{act},T})$ and $P_{\text{circ},R}(N_{\text{act},R})$ can be respectively modeled as\footnote{Note that the multiplier of $2$ applied to $P_{\text{DAC}}$ and $P_{\text{ADC}}$ explicitly accounts for the parallel In-phase (I) and Quadrature (Q) analog processing branches required in modern zero intermediate frequency architectures. In contrast, $P_{\text{mix}}$ and $P_{\text{filt}}$ are modeled as lumped parameters, which implicitly incorporate the total power consumed by the integrated I/Q modulation/demodulation module and the dual-branch filtering stage per active RF chain, respectively.}\cite{gast2024hardwareaware,abbas2017millimeter,zhang2019mixedadc}
\begin{align}
P_{\text{circ},T}(\!N_{\text{act},T}\!)\! &=\! P_{\text{LO},T}\! +\! N_{\text{act},T}\! ( 2P_{\text{DAC}}\! +\! P_{\text{LPF},T}\! +\! P_{\text{Mix},T}\!), \\
P_{\text{circ},R}(\!N_{\text{act},R}\!)\! &=\! P_{\text{LO},R}\! +\! N_{\text{act},R}\! ( 2P_{\text{ADC}}\! +\! P_{\text{LPF},R}\! +\! P_{\text{Mix},R}\! +\! P_{\text{LNA}}\!),
\end{align}
where $P_{\text{LO},T}$/$P_{\text{LO},R}$, $P_{\text{LPF},T}$/$P_{\text{LPF},R}$, and $P_{\text{Mix},T}$/$P_{\text{Mix},R}$ denote the power consumption of the transmitter-side/receiver-side local oscillator (LO), low-pass filter (LPF), and mixer, respectively, and $P_{\text{DAC}}$, $P_{\text{ADC}}$, and $P_{\text{LNA}}$ denote the power consumption of the digital-to-analog converter (DAC), analog-to-digital converter (ADC), and low noise amplifier (LNA), respectively, which can be modeled with\cite[Eq.~(9), (4), (8)]{gast2024hardwareaware}
\begin{align}
P_{\text{DAC}} &= 1.5 \cdot 10^{-5} (2^{b_{\text{DAC}}} - 1) + 9 \cdot 10^{-12} b_{\text{DAC}} \text{BW},\\
P_{\text{ADC}} &= FoM_\text{ADC} \cdot 2^{b_{\text{ADC}}} \text{BW} \sqrt{1 + \left(\frac{\text{BW}}{f_\text{cor}}\right)^2},\\
P_{\text{LNA}} &= \frac{G_{\text{LNA}} \text{BW} N_0}{(N_{\text{LNA}} - 1) FoM_{\text{LNA}}},
\end{align}
respectively, where $b_{\text{DAC}}$ and $b_{\text{ADC}}$ denote the number of DAC/ADC quantization bits, $\text{BW}$ is the system bandwidth, $FoM_\text{ADC}$ and $FoM_{\text{LNA}}$ refer to the Walden’s figure of merit (FoM) of the ADC and the LNA, respectively, $f_\text{cor}$ represents the empirically identified corner frequency, $G_{\text{LNA}}$ and $N_{\text{LNA}}$ denote the gain and noise figure of the LNA, and $N_0$ denotes the noise power spectral density.

In addition, $P_{\text{total},F}$/$P_{\text{total},G}$ combines the drive circuit power consumption $P_{\text{dc},F}(K_F)$/$P_{\text{dc},F}(K_G)$ and the Ohmic power dissipation $P_{\text{ohmic},F}(\mathbf{Y}_F)$/$P_{\text{ohmic},G}(\mathbf{Y}_G)$ stemming from the parasitic resistance of lossy TACs. Specifically, $P_{\text{dc},F}(K_F)$ and $P_{\text{dc},F}(K_G)$ are directly related to the number of TACs in the respective MiLACs (i.e., the circuit complexity $K_F$/$K_G$), given by
\begin{align}
P_{\text{dc},F}(K_F) &= K_{F} P_\text{dc,tac}, \\
P_{\text{dc},F}(K_G) &= K_{G} P_\text{dc,tac},
\end{align}
where $P_\text{dc,tac}$ denotes the drive circuit power consumption per TAC; $P_{\text{ohmic},F}(\mathbf{Y}_F)$ and $P_{\text{ohmic},G}(\mathbf{Y}_G)$ can be modeled as 
\begin{align}
P_{\text{ohmic},F}(\mathbf{Y}_F) = \mathbf{v}_F^H \Re\{\mathbf{Y}_F\} \mathbf{v}_F, \label{Pdyn_F}\\
P_{\text{ohmic},G}(\mathbf{Y}_G) = \mathbf{v}_G^H \Re\{\mathbf{Y}_G\} \mathbf{v}_G, \label{Pdyn_G}
\end{align}
using the nodal voltage analysis in Appendix \ref{App. A}, with $\mathbf{v}_F$/$\mathbf{v}_G$ denoting the nodal voltage vector across the transmitter-side/receiver-side MiLAC. According to \eqref{Pdyn_F} and \eqref{Pdyn_G}, it is worth emphasizing that $P_{\text{ohmic},F}(\mathbf{Y}_F)$/$P_{\text{ohmic},G}(\mathbf{Y}_G)$ is directly related to the conductance value of $\mathbf{Y}_F$/$\mathbf{Y}_G$.

Given $P_{\text{total}}$, the EE $\eta_{\text{EE}}$ can be defined as 
\begin{align}\label{ee}
\eta_{\text{EE}} = \frac{\text{BW} \cdot R}{P_{\text{total}}}.
\end{align}
Consequently, the EE optimization problem is formulated as
\begin{subequations}
\begin{align}
(\mathcal{P}_2): \quad & \max_{\mathbf{F}, \mathbf{G}, \mathbf{p}} 
& & \eta_{\text{EE}}\\
& \text{s.t.}
& & \sum_{s=1}^{N_s} p_s = 1, \\
&&& K_{F} \leq K_{F,\max}, ~ K_{G} \leq K_{G,\max}, \\
&&& \mathbf{F} \in \mathcal{F}, ~ \mathbf{G} \in \mathcal{G}.
\end{align}
\end{subequations}

In contrast to the SE maximization in $(\mathcal{P}_1)$, which tolerates denser receiver-side MiLAC architectures due to the noise attenuation effect, the EE optimization in $(\mathcal{P}_2)$ imposes a stringent power penalty on every active interconnection regardless of the deployment sides of the MiLACs. Since the power consumption of a TAC is fundamentally independent of its noise-attenuating properties, solving $(\mathcal{P}_2)$ inherently penalizes excessive circuit complexity along with hardware losses at both the transmitter-side and receiver-side MiLACs. Consequently, this EE-oriented formulation drives the joint design toward highly sparse and strictly energy-efficient MiLAC architectures at both the transmitter and receiver sides by retaining only those interconnections whose marginal SE contributions can strictly justify their cumulative power consumption.

To simultaneously tackle the challenges in architecture design and performance optimization in the two formulated problems, we propose the LJAPOF.

\section{Proposed Joint Architecture and Performance Optimization Framework}\label{Sec. Method}

\begin{figure*}[htbp]
\centering
\includegraphics[width=1\textwidth]{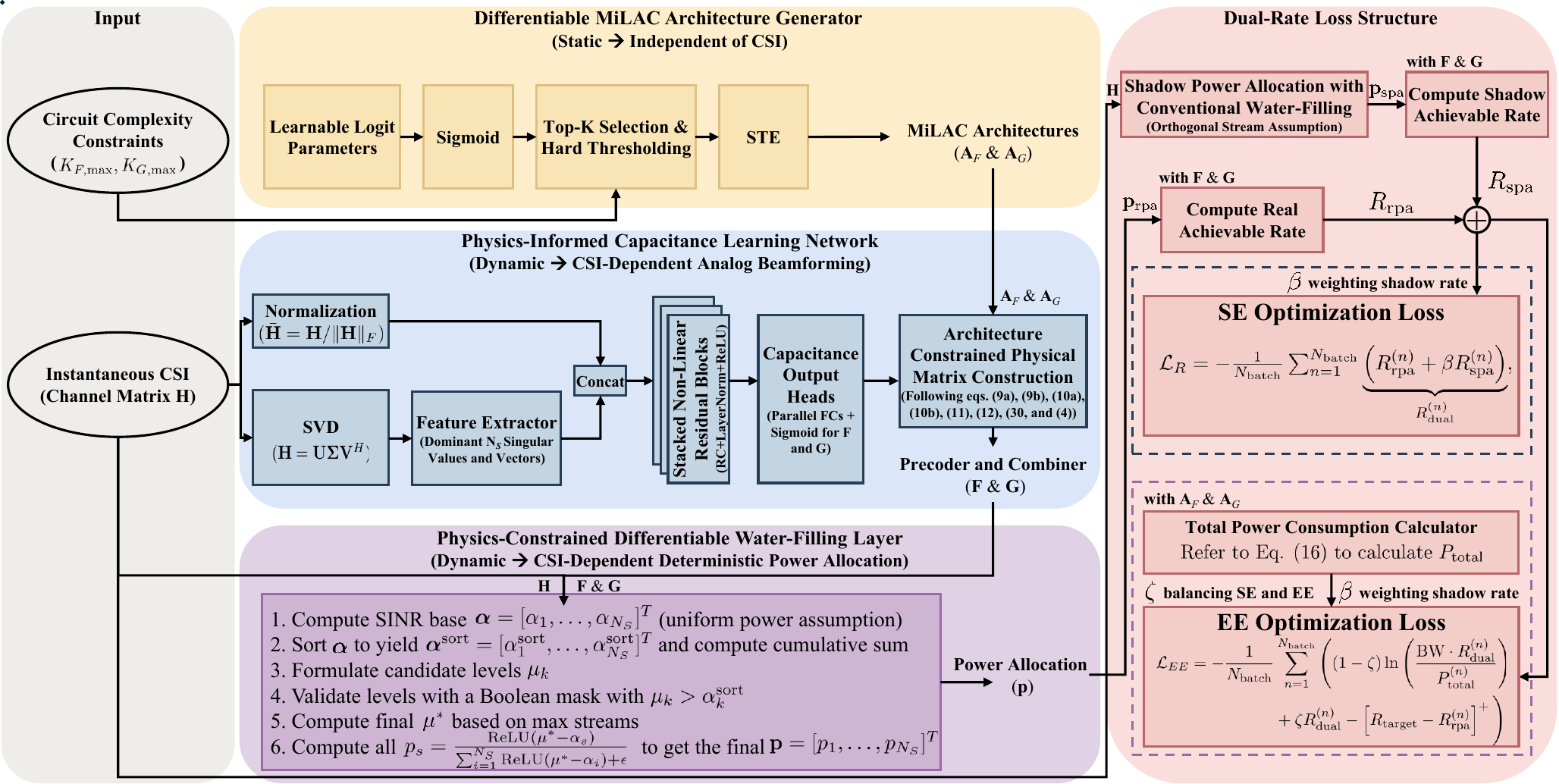}
\caption{The proposed learning-based joint architecture and performance optimization framework (LJAPOF).}
\label{framework}
\end{figure*}

As shown in Fig. \ref{framework}, the proposed LJAPOF comprises a differentiable MiLAC architecture generator for architecture design, a physics-informed capacitance learning network (PICLN) for physics-informed analog beamforming, a physics-constrained differentiable water-filling (PCDWF) layer for optimal power allocation, and a dual-rate loss structure for end-to-end unsupervised learning to optimize the system's SE or EE. Unlike conventional approaches that treat beamforming and power allocation as isolated optimization problems or rely on purely data-driven black-box models, our framework tightly integrates the physics knowledge in microwave networks and wireless communications, making it highly robust and interpretable. Note that the physical architectures of MiLACs are static, i.e., strictly independent of the instantaneous channel state information (CSI). Conversely, the physical admittance values of the activated TACs defined by the physical architectures are dynamically adapted to the instantaneous CSI to execute analog beamforming. Therefore, the differentiable MiLAC architecture generator generates an architecture for a statistical set of CSI samples, while the PICLN and the PCDWF output dynamic capacitance values and power allocation for each CSI sample.

\subsection{Differentiable MiLAC Architecture Generator}

To jointly discover the optimal architectures $\mathbf{A}_F$ and $\mathbf{A}_G$ for the lossy MiLACs at both the transmitter and the receiver, the framework must adhere to specific maximum circuit complexity constraints, denoted as $K_{F,\max}$ and $K_{G,\max}$. Since the discrete architecture characterization matrices are non-differentiable, making direct optimization intractable, we propose the differentiable MiLAC architecture generator. Note that the diagonal elements of $\mathbf{A}_F$ and $\mathbf{A}_G$ are permanently activated, while the strict lower-triangular off-diagonal elements of $\mathbf{A}_F$ and $\mathbf{A}_G$ are populated by learnable differentiable binary masks, denoted as $\mathbf{M}_F$ and $\mathbf{M}_G$, respectively. In addition, symmetry in $\mathbf{A}_F$ and $\mathbf{A}_G$ is enforced to reflect the reciprocal nature of the impedance networks. Taking the transmitter-side MiLAC as an example, $\mathbf{M}_F$ is generated as follows.

\textit{1)} Let $\boldsymbol{\alpha}_F \in \mathbb{R}^{N_F(N_F-1)/2}$ denote the learnable logits parameterizing the off-diagonal connections of the transmitter-side MiLAC. The probabilities of activating specific inter-element connections are obtained via the Sigmoid function, i.e., $\mathbf{p}_F = \sigma(\boldsymbol{\alpha}_F)$.

\textit{2)} To comply with the circuit complexity constraints $K_{F,\max}$ and $K_{G,\max}$ while removing negligible connections, we apply a hard threshold generated based on a flexible base activation mask available to dynamically prune interconnections and a budget mask to satisfy the circuit complexity constraints. Specifically, for the transmitter-side MiLAC, we define the base activation mask $\mathbf{M}_{\text{base},F} = \mathbb{I}(\mathbf{p}_F > 0.5)$ and the budget mask $\mathbf{M}_{\text{budget},F} = \mathbb{I}(\mathbf{p}_F \ge \text{topK}(\mathbf{p}_F, K_{F,\max}))$, where $\mathbb{I}(\cdot)$ is the indicator function. The exact binary architecture mask is derived as $\mathbf{M}_{\text{hard},F} = \mathbf{M}_{\text{base},F} \odot \mathbf{M}_{\text{budget},F}$, where $\odot$ denotes the Hadamard product.

\textit{3)} Since the indicator function and top-$K$ selection are non-differentiable, we employ the Straight-Through Estimator (STE) to enable end-to-end backpropagation\cite{bengio2013estimating}, with the differentiable architecture mask formulated as
\begin{equation}
\mathbf{M}_F = \mathbf{M}_{\text{hard},F} + \mathbf{p}_F - \operatorname{stopgrad}(\mathbf{p}_F),
\end{equation}
where $\operatorname{stopgrad}(\cdot)$ prevents the flow of gradients.

The differentiable architecture mask for the receiver-side MiLAC, denoted by $\mathbf{M}_G$, is generated following the identical procedure. 

\subsection{Physics-Informed Capacitance Learning Network}

While the differentiable MiLAC architecture generator supports architecture design for MiLACs, the admittance values of the activated connections are dynamically determined by the instantaneous CSI for beamforming. Given the architectures $\mathbf{A}_F$ and $\mathbf{A}_G$ generated by the architecture generator, the primary objective of the PICLN is to learn the highly non-linear mapping from the instantaneous CSI to the underlying capacitance values, which uniquely determine the admittance value of each TAC according to \eqref{R_YF} - \eqref{I_YG}. To accelerate convergence and provide a strong inductive bias, the proposed PICLN does not solely rely on the channel matrix $\mathbf{H}$. Instead, we explicitly compute the Singular Value Decomposition (SVD) of the channel, $\mathbf{H} = \mathbf{U}\mathbf{\Sigma}\mathbf{V}^H$, and extract the dominant $N_S$ singular values and their corresponding singular vectors. These SVD features are concatenated with the normalized channel matrix $\bar{\mathbf{H}} = \mathbf{H}/\Vert\mathbf{H}\Vert_F$ to form the input state $\mathbf{x}_0$, where $\Vert\cdot\Vert_F$ denotes the matrix Frobenius norm. This provides the PICLN with an explicit spatial directional guide, effectively reducing the search space for optimal admittance configuration.

A critical challenge in optimizing lossy MiLACs is the physical mapping from the capacitance values to the effective precoding and combining matrices, $\mathbf{F}$ and $\mathbf{G}$. This mapping requires constructing the admittance matrices $\mathbf{Y}_F$ and $\mathbf{Y}_G$, whose off-diagonal elements are explicitly gated by the learned architectures $\mathbf{A}_F$ and $\mathbf{A}_G$, and subsequently computing $\mathbf{F}$ and $\mathbf{G}$ via highly non-linear matrix inversions shown in \eqref{F} and \eqref{G}. When gradients are backpropagated through these complex matrix inversions, standard multi-layer perceptrons (MLPs) suffer from severe numerical instability, rendering the deep layers incapable of learning effective MiLAC capacitances.

To overcome this, we design a stacked non-linear residual block (NLRB) based deep architecture. Each of the proposed non-linear residual blocks combines a residual connection (RC) \cite{he2016deep} to facilitate gradient propagation, a layer normalization (LayerNorm) to stabilize the feature distributions\cite{ba2016layer}, and a Rectified Linear Unit (ReLU) activation function to introduce non-linearity. Let $\mathbf{x}_l$ denote the input to the $l$-th NLRB. The forward propagation through the block is formulated as
\begin{equation}
\mathbf{x}_{l+1} = \mathbf{x}_l + \text{ReLU}\big(\text{LayerNorm}(\mathbf{W}_l \mathbf{x}_l + \mathbf{b}_l)\big),
\end{equation}
where $\mathbf{W}_l$ and $\mathbf{b}_l$ are the learnable weight matrix and bias vector, respectively. Specifically, the residual skip connection $\mathbf{x}_l + \dots$ creates a gradient highway, ensuring that the micro-gradients originating from the Y-parameter matrix inversions can flow unhindered back to the initial feature extraction layers.

The output of the final NLRB is mapped through parallel fully-connected layers and passed through a Sigmoid activation function scaled by the physical hardware limits $[C_{\text{min}}, C_{\text{max}}]$ to predict the valid capacitance values for the lossy MiLACs, i.e., $C_F$ and $C_G$.

Finally, the PICLN computes the dynamic off-diagonal admittance components by multiplying the predicted capacitance values with the static architecture masks $\mathbf{M}_F$ and $\mathbf{M}_G$. These are then substituted alongside the unmasked diagonal capacities to construct the complete admittance matrices $\mathbf{Y}_F$ and $\mathbf{Y}_G$, simultaneously yielding the precoding and combining matrices $\mathbf{F}$ and $\mathbf{G}$ through exact physical modeling.

Given $C_F$ and $C_G$, the PICLN outputs the admittance values $Y_F$ and $Y_G$ according to \eqref{R_YF} - \eqref{I_YG}, generates $\mathbf{Y}_F$ and $\mathbf{Y}_G$ following \eqref{matY_F} and \eqref{matY_G}, and ultimately generates the precoder and combiner $\mathbf{F}$ and $\mathbf{G}$ based on \eqref{F} and \eqref{G}, constrained by the learned architectures $\mathbf{A}_F$ and $\mathbf{A}_G$.

\subsection{Physics-Constrained Differentiable Water-Filling Layer}

Due to the hardware losses and inter-stream interference introduced by lossy MiLACs, conventional water-filling \cite{clerckx2013mimo} is no longer optimal in lossy MiLAC-aided systems. Despite the existence of some deep learning-based power allocation methods to mitigate inter-stream interference\cite{choi2025deep, perdana2021deep, sanguinetti2018deep}, these methods are purely data-driven and rely on unconstrained parameterized modules with lots of learnable parameters, e.g., multi-layer perceptrons (MLPs). While these methods can fit well, they lack physical grounding and interpretability, introduce significant parameter overhead, and require a long learning process to be able to allocate the power properly.

To address these fundamental limitations, we propose the PCDWF layer. Rather than relying on massive learnable parameters to blindly approximate specific training distributions, we embed the exact water-filling algorithm directly into the computational graph of the proposed LJAPOF using strictly differentiable tensor operations. By doing so, the PCDWF layer deterministically computes the optimal power allocation under inter-stream interference governed by information-theoretic physics, which can be elaborated as follows.

Given the learned MiLAC matrices $\mathbf{F}$ and $\mathbf{G}$, the PCDWF layer first calculates the equivalent inverse SINR base, $\boldsymbol{\alpha} = [\alpha_1, \ldots, \alpha_{N_S}]^T$. To approximate the interference before actual power allocation, we evaluate the relative interference assuming a uniform power distribution as
\begin{equation}
\alpha_s = \frac{\frac{P_T}{N_S} \sum_{t \neq s} \vert[\mathbf{E}]_{s,t}\vert^2 + \Vert[\mathbf{G}]_{s,:}\Vert^2 \sigma^2}{P_T \vert[\mathbf{E}]_{s,s}\vert^2 + \epsilon}, \forall s = 1, \dots, N_S,
\end{equation}
where $\epsilon$ is a small constant for numerical stability. The algorithm then sorts the vector $\boldsymbol{\alpha}$ in ascending order to yield $\boldsymbol{\alpha}^{\text{sort}}=[\alpha^{\text{sort}}_1,\ldots,\alpha^{\text{sort}}_{N_S}]^T$, and computes its cumulative sum. The candidate water levels for $k \in \{1, \dots, N_S\}$ active streams are formulated as
\begin{equation}
\mu_k = \frac{1}{k} \left( 1.0 + \sum_{i=1}^{k} \alpha^{\text{sort}}_i \right).
\end{equation}
A Boolean mask validates the legitimate water levels by verifying $\mu_k > \alpha^{\text{sort}}_k$. Note that to prevent complete computational disruption caused by floating-point truncation, we enforce a hard clamp to ensure at least one active stream. The final water level $\mu^*$ is then gathered using the index of the maximum number of valid active streams.

Finally, to handle the physical truncation and enforce a hard normalization to satisfy $\sum_{s=1}^{N_S} p_s = 1$, the final power allocation vector $\mathbf{p} = [p_1, \ldots, p_{N_S}]^T$ is computed with its $s$-th element given by
\begin{equation}
p_s = \frac{\text{ReLU}(\mu^* - \alpha_s)}{\sum_{i=1}^{N_S} \text{ReLU}(\mu^* - \alpha_i) + \epsilon}.
\end{equation}

By unwrapping the iterative algorithm into a continuous computational graph, the differentiability of this layer is mathematically guaranteed. Operations such as sorting and gathering safely route the gradients, acting as a strict physical constraint. Crucially, the global water level $\mu^*$ remains a function of the interference levels of the active streams, guaranteeing that the power allocation is always physically optimal for the current hardware state and forcing it to adjust the MiLAC capacitances to suppress inter-stream interference.

Compared with existing deep learning-based power allocation methods, this parameter-free and physics-constrained design not only reduces the computational complexity of the LJAPOF but also ensures strict white-box interpretability and robust cross-scenario generalization. While compared with the conventional water-filling algorithm \cite{clerckx2013mimo}, which is a static and post-processing routine designed for perfectly orthogonal channels, our PCDWF layer is dynamic and interference-aware, actively compensating for hardware-induced losses and adaptively achieving optimal power allocation across mutually interfering streams.

\subsection{SE Optimization with the Dual-Rate Loss Structure}

We first employ the proposed optimization framework to optimize the system's SE in an unsupervised learning manner. While the PCDWF layer guarantees a physically optimal power allocation, utilizing only its output to compute the achievable rate as the loss function introduces a catastrophic combination of shortcut learning \cite{geirhos2020shortcut} and gradient vanishing. In the early stage of optimization, the unoptimized MiLACs inherently generate severe inter-stream interference. Following strict physical laws, the PCDWF layer naturally exploits a suboptimal shortcut: rather than optimizing the MiLAC capacitances for interference suppression, they assign zero power to the heavily interfered streams to artificially boost the sum rate of the remaining streams, which creates a severe optimization trap: setting a stream's power to zero inherently cuts off its gradient flow during backpropagation, making the PICLN permanently lose the ability to update the precoding and combining matrices to suppress that interference and trapping the system in a severely degraded single-stream local optimum.

To fully overcome this shortcut learning and gradient vanishing bottleneck, we design a dual-rate loss structure. Alongside the real power allocation obtained by the PCDWF layer, we introduce a parallel and purely virtual ``shadow power allocation'' obtained by the conventional water-filling algorithm, as presented in \eqref{water-filling solution}. By leveraging the conventional water-filling algorithm, which achieves optimal power allocation under the assumption of independent and orthogonal streams, the shadow power allocation guides the PICLN to maintain an appropriate number of data streams and thus prevent the single-stream trap.

Specifically, during the forward pass, for a specific channel realization $\mathbf{H}^{(n)}$, where $n \in \{1, \ldots, N_{\text{batch}}\}$ with $N_{\text{batch}}$ denoting the total number of channel realizations in a batch, the PICLN learns the corresponding valid capacitances $C_F^{(n)}$ and $C_G^{(n)}$, which are deterministically mapped to the analog precoding matrix $\mathbf{F}^{(n)}$ and combining matrix $\mathbf{G}^{(n)}$.
To generalize the performance metric, we define the SINR for the $s$-th data stream and the $n$-th channel sample in the batch as a function of a given power allocation vector $\mathbf{p}^{(n)} = [p_1^{(n)}, \ldots, p_{N_S}^{(n)}]^T$ as
\begin{equation}
\text{SINR}_s^{(n)}(\mathbf{p}^{(n)}) = \frac{P_T p_s^{(n)} \left\vert [\mathbf{E}^{(n)}]_{s,s} \right\vert^2}{P_T \sum_{t \neq s} p_t^{(n)} \left\vert [\mathbf{E}^{(n)}]_{s,t} \right\vert^2 + \left\Vert [\mathbf{G}^{(n)}]_{s,:} \right\Vert^2 \sigma^2}, \label{sinr_ch}
\end{equation}
where $\mathbf{E}^{(n)} = \mathbf{G}^{(n)}\mathbf{H}^{(n)}\mathbf{F}^{(n)}$ denotes the effective channel. 
Let $\mathbf{p}_{\text{rpa}}^{(n)}$ and $\mathbf{p}_{\text{spa}}^{(n)} = [p_1^{\star(n)}, \ldots, p_{N_S}^{\star(n)}]^T$ denote the real power allocation obtained by the PCDWF layer and the shadow power allocation obtained by the conventional water-filling algorithm, respectively. We then compute two distinct achievable rate metrics as
\begin{align}
R_{\text{rpa}}^{(n)} &= \sum_{s=1}^{N_S} \log_2 \left( 1 + \text{SINR}_s^{(n)}(\mathbf{p}_{\text{rpa}}^{(n)}) \right), \\
R_{\text{spa}}^{(n)} &= \sum_{s=1}^{N_S} \log_2 \left( 1 + \text{SINR}_s^{(n)}(\mathbf{p}_{\text{spa}}^{(n)}) \right).
\end{align}
The final loss function $\mathcal{L}_{R}$ for SE optimization is formulated as the negative empirical average of the sum of both rates:
\begin{equation}
\mathcal{L}_{R} = - \frac{1}{N_{\text{batch}}} \sum_{n=1}^{N_{\text{batch}}} \underbrace{\left( R_{\text{rpa}}^{(n)} + \beta R_{\text{spa}}^{(n)}\right)}_{R_{\text{dual}}^{(n)}}, \label{loss_dual}
\end{equation}
where $\beta$ is a hyperparameter to allow flexible weighting of $R_{\text{spa}}^{(n)}$ relative to $R_{\text{rpa}}^{(n)}$ during the optimization. Note that $\beta$ is initialized with $\beta_0$ and then gradually reduced to $0$, so that the final objective of \eqref{loss_dual} is consistent with $(\mathcal{P}_1)$.

This tailored dual-rate loss function $\mathcal{L}_{R}$ is highly synergistic with the optimization problem. The shadow achievable rate $R_{\text{spa}}^{(n)}$ acts as a continuous gradient-preserving anchor to avoid the single-stream trap. Specifically, since $\mathbf{p}_{\text{spa}}$ guarantees appropriate power allocation for data streams regardless of the noise levels, the denominator of \eqref{sinr_ch} constantly exposes the PICLN to the hardware loss-induced inter-stream interference of all spatial dimensions. During backpropagation, the gradients from $R_{\text{spa}}^{(n)}$ flow seamlessly backward, mathematically penalizing the off-diagonal elements $[\mathbf{E}]_{s,t}$ (i.e., inter-stream interference) when the MiLACs are not fully optimized at the very beginning of optimization. Concurrently, the real achievable rate $R_{\text{rpa}}^{(n)}$ drives the PICLN to push the absolute physical capacity limits of the system. This continuous dual-path gradient flow fundamentally eliminates shortcut learning by enforcing the upstream PICLN to continuously adjust the admittance components of the MiLACs to mitigate inter-stream interference, maximize the spatial multiplexing gain, and ensure robust convergence.

\subsection{EE Optimization with the Dual-Rate Loss Structure}

Similar to the SE optimization, naively formulating the loss function for EE optimization as the negative empirical average of EE under various channel realizations encounters the previously discussed shortcut learning problem. To circumvent this severe performance collapse, we embed the dual-rate structure into the EE objective. In addition, although $(\mathcal{P}_2)$ aims to maximize the system's EE, practical implementations must carefully navigate the fundamental tradeoff between EE and SE. This balance is crucial to prevent the LJAPOF from indiscriminately disabling MiLAC interconnections to minimize total power consumption, which is an aggressive power-saving measure that would cause severe SE degradation and compromise the system's quality of service (QoS). To establish this balance, we introduce a hybrid objective. Specifically, we utilize a weighting parameter $\zeta$ to seamlessly balance the EE and SE with the dual-rate structure embedded, while concurrently incorporating an SE penalty term to strictly enforce the baseline QoS. By leveraging the gradient-preserving anchor provided by the shadow achievable rate $R_{\text{spa}}^{(n)}$, the final dual-rate EE loss function, $\mathcal{L}_{EE}$, is formulated as
\begin{equation}
\begin{aligned}
\mathcal{L}_{EE} = - \frac{1}{N_{\text{batch}}} &\sum_{n=1}^{N_{\text{batch}}} \Bigg(  (1 - \zeta)\ln\left(\frac{\text{BW} \cdot R_{\text{dual}}^{(n)}}{P_{\text{total}}^{(n)}}\right) \\
& + \zeta R_{\text{dual}}^{(n)}  - \left[ R_{\text{target}} - R_{\text{rpa}}^{(n)} \right]^+ \Bigg),
\end{aligned} \label{loss_ee_dual}
\end{equation}
where $\zeta \in [0, 1)$ denotes the SE weighting parameter, $R_{\text{target}}$ denotes the target SE, $[x]^+ \triangleq \max(0, x)$, and $\ln(\cdot)$ denotes the natural logarithm adopted to stabilize the training gradients.

By incorporating the dual-rate structure, $\mathcal{L}_{EE}$ establishes a robust and dynamic optimization process. On the one hand, the total power consumption $P_{\text{total}}^{(n)}$ imposes a continuous sparsity-inducing pressure, forcing the PICLN to identify and deactivate redundant or highly lossy TACs. On the other hand, the shadow achievable rate $R_{\text{spa}}^{(n)}$ preserves the gradients for necessary spatial dimensions, acting as a strict penalty against rank collapse and single-stream shortcuts. Consequently, the PICLN is compelled to intelligently prune the MiLAC architectures without sacrificing spatial multiplexing, ultimately achieving a highly sparse and energy-efficient architecture that maintains a superior SE.

\section{Simulations} \label{Sec. Sim.}
\subsection{Simulation Setup}
\subsubsection{System Setup}
To evaluate the performance of the proposed LJAPOF, a point-to-point MIMO system with both the transmitter and the receiver configured with $N_T = N_R = 32$ antennas and $N_S \in \{4, 8, 12, 16\}$ streams is considered. The signal frequency $f$ is set to $2.4 \text{ GHz}$ and the system bandwidth is $\text{BW}=100 \text{ MHz}$. We set $\eta=52.1\%$ to reflect a typical power-added efficiency to align with practical sub-6 GHz RF front-end implementations, benchmarked against the NXP A3G26D055N Airfast RF power amplifier. The total transmit power $P_T$ is $20\text{ dBm}$. For parameters related to the power consumption of the circuits in the transmitter and receiver, we set $P_{\text{LO},T}=P_{\text{LO},R}=6\text{ mW}$\cite{gast2024hardwareaware}, $P_{\text{LPF},T}=P_{\text{LPF},R}=2.5\text{ mW}$\cite{zhang2019mixedadc}, $P_{\text{Mix},T}=P_{\text{Mix},R}=1.57\text{ mW}$\cite{gast2024hardwareaware},  $b_{\text{DAC}}=4$, $b_{\text{ADC}}=4$ with $FoM_\text{ADC}=494 \text{ fJ/step/Hz}$\cite{chung200975gs} and $f_\text{cor} = 560\text{ MHz}$\cite{gast2024hardwareaware}, 
$G_{\text{LNA}}=15\text{ dB}$\cite{gast2024hardwareaware}, $N_{\text{LNA}} = 5\text{ dB}$\cite{gast2024hardwareaware}, $FoM_\text{LNA}=10^{-9}$\cite{mezghani2011modeling}, and $N_0=-174 \text{ dBm/Hz}$. Following the settings in \cite{peng2026lossy} with the varactor diode SMV2020-079LF employed to model the TAC, we set $L_1 = 6\text{ nH}$, $L_2 = 0.7\text{ nH}$, $R_1=1~\Omega$, $[C_F]_{\text{min}} = [C_G]_{\text{min}} = 0.35\text{ pF}$, and $[C_F]_{\text{max}} = [C_G]_{\text{max}} = 3.20\text{ pF}$. The reference admittance $Y_0$ is set to $1/50\text{ S}$. In addition, \cite{wang2024reconfigurable} reported a total drive circuit power consumption of $1720 \text{ mW}$ for $128$ varactor-diode-based RIS elements. Consequently, we set $P_\text{dc,tac}=1720/128=13.4375 \text{ mW}$ as the drive circuit power consumption per TAC implemented based on the varactor diode. The noise power $\sigma^2$ is set according to the SNR $\gamma = P_T/\sigma^2$, where $\gamma$ is set to $0\text{ dB}$ in this work. All results are averaged over $100$ independent and identically distributed (i.i.d.) Rayleigh fading channel realizations.

\subsubsection{Hyper-parameter Settings}
The proposed framework is implemented using PyTorch\cite{paszke2019pytorch}. We apply a maximum of $30,000$ iterations and an early stopping strategy with $4,000$ iterations for joint architecture and performance optimization. The batch size is set to $N_{\text{batch}}=100$. We adopt the Adam optimizer with a learning rate of $0.001$. The shadow rate weighting hyper-parameter $\beta$ in \eqref{loss_dual} and \eqref{loss_ee_dual} is initialized with $\beta_0=1$ and linearly decreased to $0$ in $20,000$ epochs. The SE weighting parameter $\zeta$ in \eqref{loss_ee_dual} is set to $0.2$. The target SE $R_{\text{target}}$ in \eqref{loss_ee_dual} for EE optimization is set to the optimal SE that can be achieved in SE optimization, whereas this can be flexibly set based on practical SE QoS requirements. Regarding the deep learning architecture, the feature extraction and mapping rely on a cascade of $3$ NLRBs, each configured with a hidden dimension of $768$ neurons to ensure sufficient representational capacity for the highly non-linear Y-parameter to effective channel mappings.

The detailed system setup and hyper-parameter configurations are summarized in Table \ref{tab:parameters}.

\begin{table}[htbp]
\centering
\caption{System Setup and Hyper-parameter Configurations}
\label{tab:parameters}
\begin{tabular}{l c}
\toprule
\textbf{Parameter} & \textbf{Value} \\
\midrule
\multicolumn{2}{c}{\textit{System Setup}} \\
\midrule
No. of Tx and Rx Antennas $N_T=N_R$ & $32$ \\
No. of Data Streams $N_S$ & $\{4, 8, 12, 16\}$ \\
Signal Frequency $f$ & $2.4\text{ GHz}$ \\
System Bandwidth $\text{BW}$ & $100 \text{ MHz}$ \\
Power Added Efficiency $\eta$ & $52.1\%$ \\
Total Transmit Power $P_T$ &  $20\text{ dBm}$ \\
Power Consumption of LOs $P_{\text{LO},T}=P_{\text{LO},R}$ & $6\text{ mW}$\cite{gast2024hardwareaware}\\
Power Consumption of LPFs $P_{\text{LPF},T}=P_{\text{LPF},R}$ & $2.5\text{ mW}$\cite{zhang2019mixedadc} \\
Power Consumption of Mixers $P_{\text{Mix},T}=P_{\text{Mix},R}$ &  $1.57\text{ mW}$\cite{gast2024hardwareaware}\\
No. of DAC/ADC Quantization Bits $b_{\text{DAC}}$, $b_{\text{ADC}}$ &  $4$,~$4$ \\
FoM of the ADC $FoM_\text{ADC}$ & $494 \text{ fJ/step/Hz}$\cite{chung200975gs} \\
Corner Frequency $f_\text{cor}$ & $560\text{ MHz}$\cite{gast2024hardwareaware} \\
LNA Gain $G_{\text{LNA}}$ & $15\text{ dB}$\cite{gast2024hardwareaware} \\
LNA Noise Figure $N_{\text{LNA}}$ & $5\text{ dB}$\cite{gast2024hardwareaware} \\
FoM of the LNA $FoM_\text{LNA}$ & $10^{-9}$\cite{mezghani2011modeling} \\
Inductances $L_1$, $L_2$ & $6\text{ nH}$, $0.7\text{ nH}$ \\
Parasitic Resistance $R$ & $1\ \Omega$ \\
Minimum Capacitance $[C_F]_{\text{min}} = [C_G]_{\text{min}}$ & $0.35\text{ pF}$ \\
Maximum Capacitance $[C_F]_{\text{max}} = [C_G]_{\text{max}}$ & $3.20\text{ pF}$ \\
Drive Circuit Power Consumption Per TAC $P_\text{dc,tac}$ & $13.4375\text{ mW}$\\
\midrule
\multicolumn{2}{c}{\textit{Training \& Network Hyper-parameters}} \\
\midrule
Training Iterations & $30,000$ \\
Early Stopping Iterations & $4,000$ \\
Batch Size & $100$ \\
Learning Rate & $0.001$ \\
Initial Shadow Rate Weighting Factor $\beta_0$ & $1$ \\
SE Weighting Factor $\zeta$ & $0.2$ \\
Number of NLRBs & $3$ \\
Hidden Layer Dimension & $768$ \\
\bottomrule
\end{tabular}
\end{table}

\subsection{Effectiveness of the LJAPOF on SE Optimization Under Fixed MiLAC Architectures}

\begin{figure}[t]
\centering
\includegraphics[width=0.8\columnwidth]{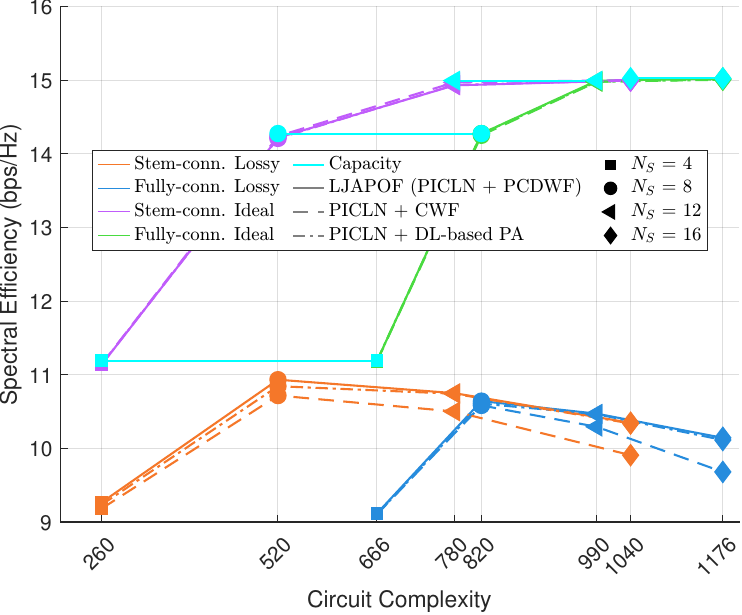}
\caption{SE of ideal and lossy MiLAC-aided MIMO systems versus circuit complexity across different fixed MiLAC architectures, numbers of streams $N_S$, and power allocation algorithms (i.e., the PCDWF layer in the proposed LJAPOF, the conventional water-filling algorithm\cite{clerckx2013mimo}, and the deep-learning based power allocation algorithm\cite{perdana2021deep}).}
\label{fig:lossy-32-power-allocation}
\end{figure}

Prior to exploiting the proposed LJAPOF for joint architecture design and performance optimization, we verify the effectiveness of the LJAPOF on SE optimization under fixed MiLAC architectures.\footnote{Note that in this subsection, the transmitter-side and receiver-side MiLACs are set to the same architecture.} First, we apply the LJAPOF to ideal MiLAC-aided MIMO systems for SE optimization while fixing the MiLAC architectures to fully-connected/stem-connected MiLACs, as the system's SE achieved with these MiLAC architectures has been validated in \cite{nerini2026mimo}. As shown in Fig. \ref{fig:lossy-32-power-allocation}, with the proposed LJAPOF, both ideal fully-connected and stem-connected MiLACs achieve the SE close to the system capacity across different numbers of streams, with neglected performance gaps arising from the small approximation errors of neural networks in approximating the global optimal solutions with a batch of channel realizations and finite learnable parameters. In addition, as $N_S$ increases, the SE that can be achieved by ideal fully-connected and stem-connected MiLACs initially improves progressively and subsequently tends to stabilize, exhibiting a clear diminishing marginal effect. These results align perfectly with the conclusions in \cite{nerini2026mimo} and demonstrate the effectiveness of the LJAPOF.

Then, by applying the proposed LJAPOF to fully-connected and stem-connected lossy MiLAC-aided MIMO systems for SE optimization, we find that stem-connected lossy MiLACs always outperform fully-connected lossy MiLACs given the same $N_S$ setting, as stem-connected MiLACs have lower circuit complexity (i.e, $K_F^\text{Stem} = N_S(2N_T+1)$ and $ K_G^\text{Stem}=N_S(2N_R+1)$) compared with fully-connected MiLACs (i.e., $K_F^\text{Fully} = (N_S+N_T)(N_S+N_T+1)/2$, $ K_G^\text{Fully}=(N_S+N_R)(N_S+N_R+1)/2$), which thus incur less hardware losses and SE degradation. Additionally, the SE achieved by lossy fully-connected and stem-connected MiLACs exhibits an initial improvement before ultimately declining with the increasing $N_S$. This behavior can be attributed to the interplay between the multiplexing gains and hardware losses of lossy MiLACs: while increasing $N_S$ yields diminishing marginal multiplexing gains, the hardware losses induced by the increased circuit complexity (i.e., $K_F^\text{Stem}$, $K_G^\text{Stem}$, $K_F^\text{Fully}$, and $K_G^\text{Fully}$) increase concurrently, ultimately overshadowing the marginal multiplexing gains and thus leading to an overall SE degradation in the high-$N_S$ regime.

Moreover, Fig. \ref{fig:lossy-32-power-allocation} also verifies the effectiveness of the PCDWF layer in the proposed LJAPOF under hardware losses and inter-stream interference compared with existing power allocation algorithms. We replace the power allocation algorithm in the proposed LJAPOF with the conventional water-filling algorithm\cite{clerckx2013mimo} and the deep-learning based power allocation algorithm\cite{perdana2021deep}, yielding the curves labeled by ``PICLN + CWF'' and ``PICLN + DL-based PA'', respectively. Notably, all three power allocation algorithms achieve near-capacity SE when applied to ideal MiLACs. For lossy MiLAC cases, both the PCDWF layer and the deep-learning-based power allocation algorithm outperform the conventional water-filling algorithm, showing their effectiveness in allocating power properly under inter-stream interference. The proposed PCDWF layer achieves an SE comparable to that of the deep-learning-based power allocation algorithm. However, unlike the deep-learning-based algorithm, which necessitates a massive overhead of $315,268$ to $513,424$ parameters (for $N_S \in [4, 16]$) and prolonged training times to map channel realizations to power allocations, the PCDWF layer is completely parameter-free and intrinsically provides clear physical interpretability.

\begin{figure}[t]
\centering
\includegraphics[width=0.8\columnwidth]{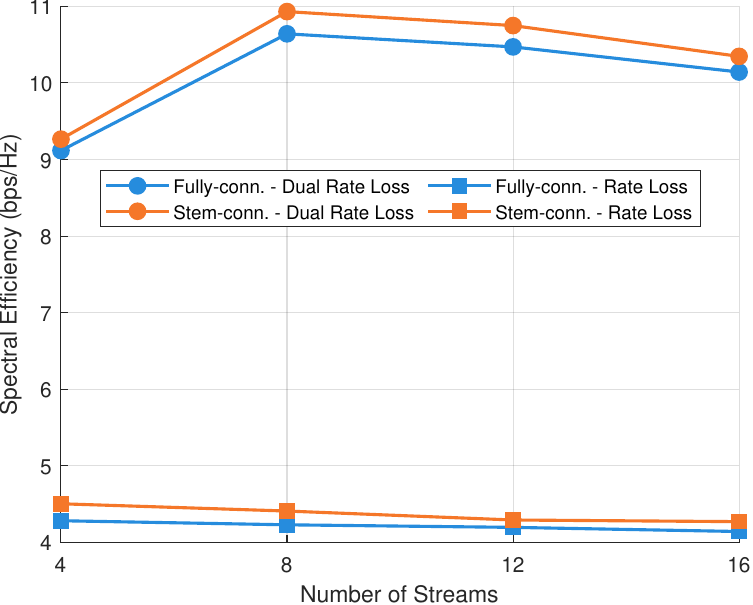}
\caption{SE of lossy MiLAC-aided MIMO systems versus the number of streams $N_S$ when optimized with $\mathcal{L}_{R}$ or $\widetilde{\mathcal{L}}_{R}$.}
\label{fig:lossy-32-dual-rate}
\end{figure}

\begin{figure}[htbp]
\centering
\includegraphics[width=0.8\columnwidth]{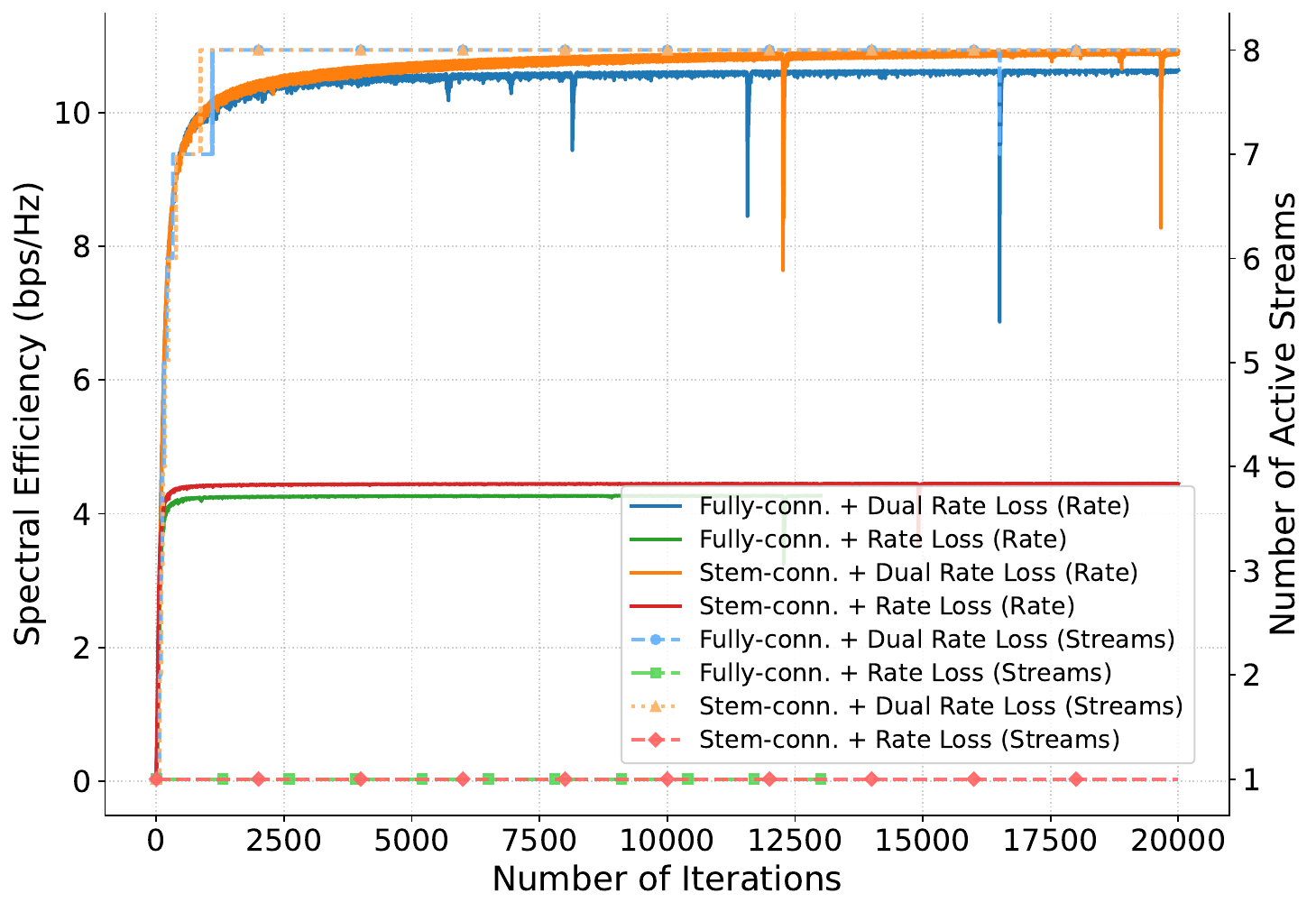}
\caption{The SE and the number of active streams of lossy MiLAC-aided MIMO systems with $N_S=8$ during the optimization process when optimized with $\mathcal{L}_{R}$ or $\widetilde{\mathcal{L}}_{R}$.}
\label{fig:lossy-32-dual-rate-iteration-curve}
\end{figure}

We further evaluate the effectiveness of the proposed dual-rate loss structure. As a comparison, we also apply purely the real achievable rate loss to maximize the SE, i.e., $\widetilde{\mathcal{L}}_{R} = - \frac{1}{N_{\text{batch}}} \sum_{n=1}^{N_{\text{batch}}} R_{\text{rpa}}^{(n)}$. As shown in Fig. \ref{fig:lossy-32-dual-rate}, both the fully-connected and stem-connected MiLACs can only yield low SE across different $N_S$ settings when optimized with $\widetilde{\mathcal{L}}_{R}$, as the LJAPOF falls into the single-stream trap. As such, with the increasing $N_S$, the system's SE drops continuously due to the escalating hardware losses. In contrast, the dual-rate loss $\mathcal{L}_{R}$ avoids shortcut learning and the single-stream trap, allowing the system to benefit from more active streams to achieve better SE. Taking $N_S=8$ as an example, the system's SE and the number of active streams during the optimization process are shown in Fig. \ref{fig:lossy-32-dual-rate-iteration-curve}. When applying $\widetilde{\mathcal{L}}_{R}$, the number of active streams remains fixed at $1$ from the beginning throughout the entire optimization process, resulting in consistently low SE. In contrast, when applying $\mathcal{L}_{R}$, the number of active streams gradually increases, ultimately maximizing the system’s SE.

\subsection{SE and EE of Lossy MiLAC-aided MIMO Systems Under Joint Architecture and Performance Optimization}\label{Sec. Sim. SE_EE}

\begin{figure*}[t]
    \centering
    \begin{subfigure}[b]{0.82\columnwidth}
        \centering
        \includegraphics[width=\linewidth]{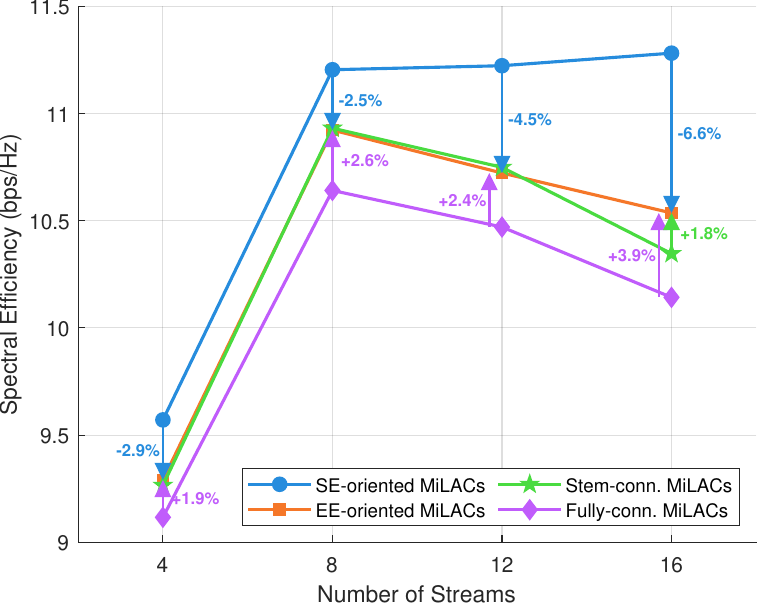}
        \caption{SE of lossy MiLAC-aided MIMO systems.}
        \label{fig:lossy32-SE-NS}
    \end{subfigure}
    \hfill
    \begin{subfigure}[b]{0.82\columnwidth}
        \centering
        \includegraphics[width=\linewidth]{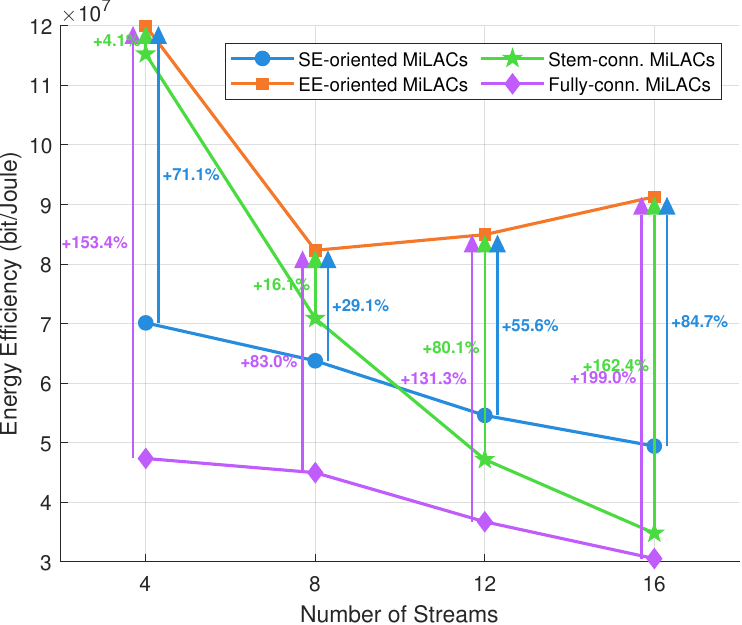}
        \caption{EE of lossy MiLAC-aided MIMO systems.}
        \label{fig:lossy32-EE-NS}
    \end{subfigure}
    \hfill
    \begin{subfigure}[b]{0.82\columnwidth}
    \centering
    \includegraphics[width=\linewidth]{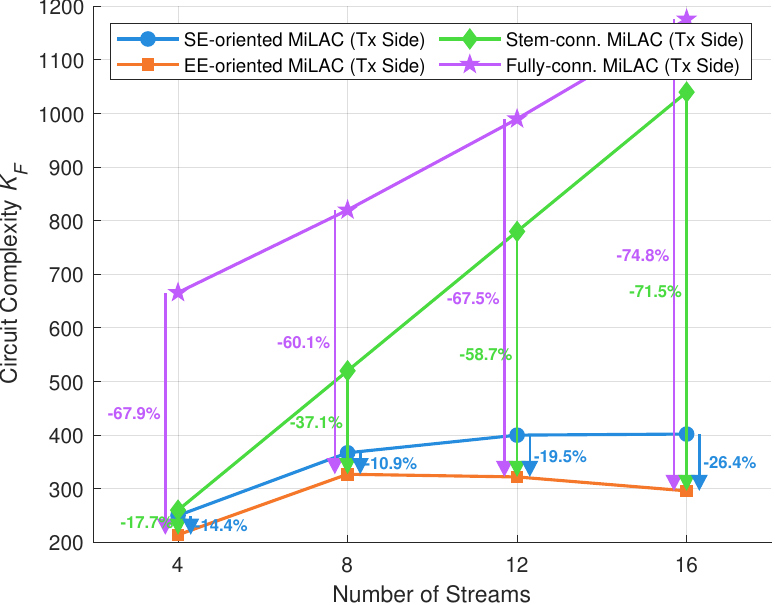}
    \caption{Circuit complexity of the transmitter-side MiLAC.}
    \label{fig:lossy32-CCF-NS}
    \end{subfigure}
    \hfill
    \begin{subfigure}[b]{0.82\columnwidth}
        \centering
        \includegraphics[width=\linewidth]{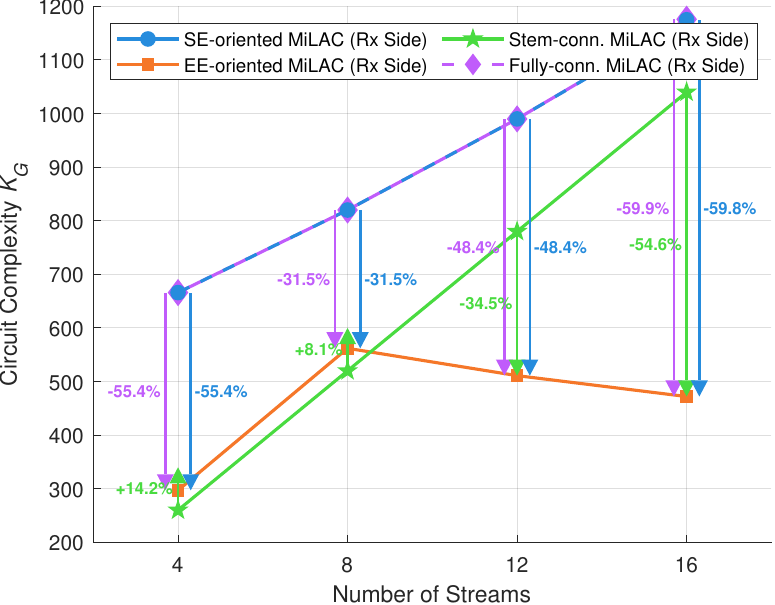}
        \caption{Circuit complexity of the receiver-side MiLAC.}
        \label{fig:lossy32-CCG-NS}
    \end{subfigure}

    \caption{SE and EE of lossy MiLAC-aided MIMO systems and the circuit complexity of the transmitter-side and receiver-side MiLACs versus the number of streams $N_S$ under joint architecture and performance optimization.}
    \label{fig:lossy32-SE-EE-CCF-CCG-NS}
\end{figure*}

After evaluating the effectiveness of the proposed LJAPOF under fixed MiLAC architectures, we investigate architecture design for lossy MiLACs and the SE and EE of lossy MiLAC-aided MIMO systems using our LJAPOF. Specifically, the transmitter-side MiLAC architecture $\mathbf{A}_F$, the receiver-side MiLAC architecture $\mathbf{A}_G$, their corresponding analog precoder $\mathbf{F}$ and $\mathbf{G}$, and power allocation ratios are learned simultaneously to optimize the SE and EE of the system via \eqref{loss_dual} and \eqref{loss_ee_dual}, respectively. Fig. \ref{fig:lossy32-SE-EE-CCF-CCG-NS} presents the SE and EE of lossy MiLAC-aided MIMO systems, along with the circuit complexity of the transmitter-side and receiver-side MiLACs, versus the number of streams $N_S$ under joint architecture and performance optimization. For clarity, we use ``SE-oriented MiLACs'' and ``EE-oriented MiLACs'' to denote the transmitter-side and receiver-side MiLACs obtained via SE and EE optimization, respectively, in these figures. For comparison, baseline schemes deploying purely stem-connected/fully-connected MiLACs at both the transmitter and receiver sides are denoted by ``Stem-conn. MiLACs'' and ``Fully-conn. MiLACs'', respectively.

\begin{figure*}[t]
    \centering
    \begin{subfigure}[b]{0.65\columnwidth}
        \centering
        \includegraphics[width=\linewidth]{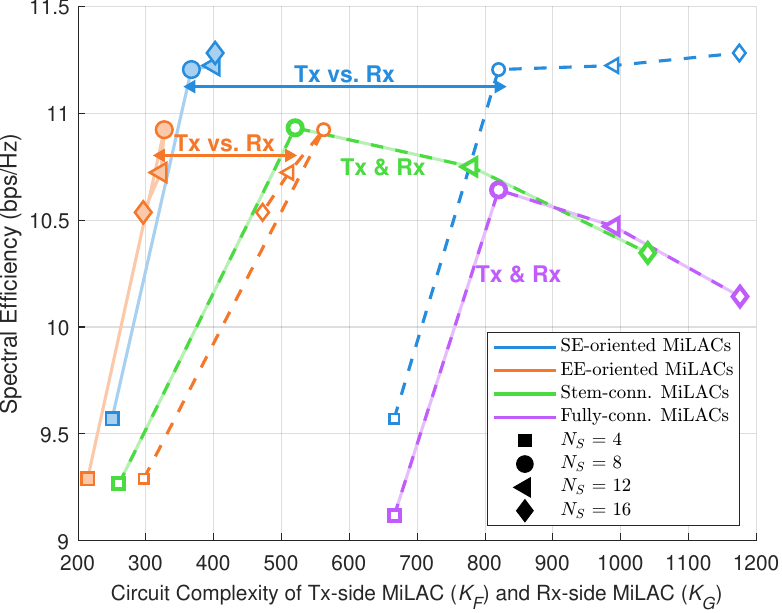}
        \caption{SE-circuit complexity tradeoff in lossy MiLAC-aided MIMO systems.}
        \label{fig:lossy32-SE-CCF-CCG}
    \end{subfigure}
    \hfill
    \begin{subfigure}[b]{0.65\columnwidth}
        \centering
        \includegraphics[width=\linewidth]{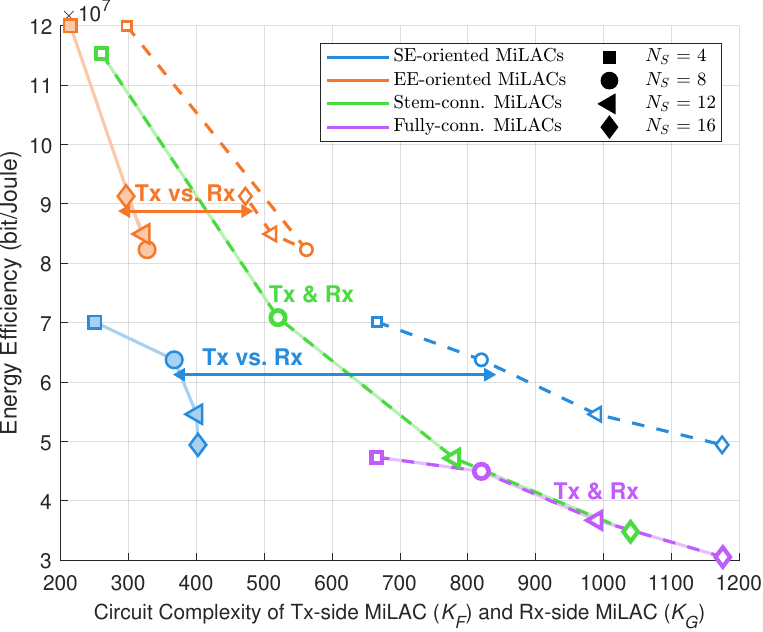}
        \caption{EE-circuit complexity tradeoff in lossy MiLAC-aided MIMO systems.}
        \label{fig:lossy32-EE-CCF-CCG}
    \end{subfigure}
    \hfill
    \begin{subfigure}[b]{0.65\columnwidth}
    \centering
    \includegraphics[width=\linewidth]{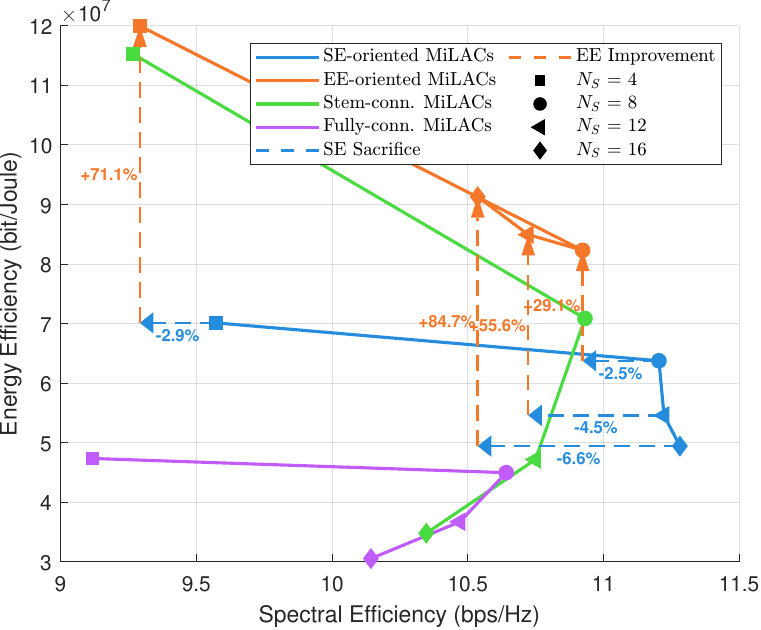}
    \caption{EE-SE tradeoff in lossy MiLAC-aided MIMO systems.}
    \label{fig:lossy32-Tradeoff-SE-EE}
    \end{subfigure}
    
    \caption{(a) SE--circuit complexity, (b) EE-circuit complexity, and (c) EE-SE tradeoffs in lossy MiLAC-aided MIMO systems.}
    \label{fig:lossy32-SE-EE-CCF-CCG-Tradeoff}
\end{figure*}

As shown in Fig. \ref{fig:lossy32-SE-NS}, the SE-oriented MiLACs achieve the highest SE compared with the EE-oriented, stem-connected, and fully-connected MiLACs under the same $N_S$ setting. Different from stem-connected and fully-connected MiLAC-aided systems, whose SE first increases and then decreases with the increasing $N_S$, the SE of the SE-oriented MiLAC-aided system first increases and then stabilizes. This is because the LJAPOF intelligently prunes the interconnections of the transmitter-side MiLAC (see Fig. \ref{fig:lossy32-CCF-NS}) to mitigate its hardware losses while maximizing the SE. It is worth emphasizing that via SE optimization, the transmitter-side MiLAC becomes sparse-connected, while the receiver-side MiLAC remains fully-connected (see Fig. \ref{fig:lossy32-CCG-NS}), exhibiting asymmetry in interconnection sparsity.\footnote{Note that the superiority of stem-connected MiLACs over fully-connected MiLACs in Fig. \ref{fig:lossy-32-power-allocation} does not contradict the optimality of a fully-connected receiver-side MiLAC. Since the ``Fully-connected Lossy'' baseline is defined to deploy fully-connected MiLACs at both the transmitter and receiver sides, the massive hardware losses originating from the dense transmitter-side MiLAC effectively negate any multiplexing benefits gained at the receiver, ultimately degrading the net SE.} This is because the combining is natively executed in the analog domain by the receiver-side MiLAC, such that the hardware losses in the combiner $\mathbf{G}$ attenuate the noise alongside the signal, as reflected in \eqref{sinr}. In contrast, the hardware losses in the transmitter-side MiLAC only attenuate the signal, making it more sensitive to the SE. This mathematical asymmetry naturally causes the transmitter-side MiLAC to converge to a sparse architecture while the receiver-side MiLAC remains fully-connected. In addition, the SE-oriented MiLACs achieve $3.3\%$ and $2.5\%$ improvements in SE compared with the stem-connected MiLACs for $N_S=4$ and $N_S=8$, at the cost of $39.2\%$ and $10.0\%$ of EE (see Fig. \ref{fig:lossy32-EE-NS}), respectively, making it less attractive. While in the high $N_S$ regime with $N_S=12$ and $N_S=16$, the SE-oriented MiLACs attain $4.4\%$ and $9.0\%$ gains in SE, along with $15.7\%$ and $42.1\%$ improvements in EE, respectively, thanks to the significantly reduced circuit complexity of the transmitter-side MiLAC, despite the receiver-side MiLAC remaining fully-connected.

As shown in Fig. \ref{fig:lossy32-EE-NS}, the EE-oriented MiLACs achieve the highest EE among these architectures under the same $N_S$ setting. Specifically, the EE-oriented MiLACs and stem-connected MiLACs achieve comparable SE for $N_S=4$ and $N_S=8$, while the former ones achieve $4.1\%$ and $16.1\%$ higher EE, respectively. These EE gains are achieved by reducing the transmitter-side MiLAC circuit complexity by $17.7\%$ and $37.1\%$, respectively, even though the receiver-side MiLAC has $14.2\%$ and $8.1\%$ higher circuit complexity, respectively. While for $N_S=12$ and $N_S=16$, the EE-oriented MiLACs significantly outperform stem-connected MiLACs by $80.1\%$ and $162.4\%$ higher EE, respectively, along with comparable or slightly higher SE. These significant EE improvements arise from the reduced circuit complexity of both the transmitter-side and receiver-side MiLACs (by at most $71.5\%$ for the transmitter-side MiLAC and $54.6\%$ for the receiver-side MiLAC, respectively). Moreover, Fig. \ref{fig:lossy32-CCF-NS} and \ref{fig:lossy32-CCG-NS} again show asymmetrically sparse transmitter-side and receiver-side MiLAC architectures when obtained via EE optimization. This reveals that even though via EE optimization, the receiver-side MiLAC becomes sparse-interconnected but always maintains denser than the transmitter-side MiLAC. This is because the LJAPOF finds that activating more interconnections in the receiver-side MiLAC to some extent helps increase the spatial multiplexing gains and EE at the cost of a bit of additional power consumption, as a denser receiver-side MiLAC enables better harvesting of spatial energy leaked due to hardware losses and consequent inter-stream interference.

In addition, it is worth noting that deploying fully-connected MiLACs simultaneously at both the transmitter and receiver sides consistently yields the poorest SE and EE while incurring the highest circuit complexity across all evaluated scenarios. This demonstrates that such dense architectures are fundamentally ill-suited for practical MIMO deployments, as the anticipated spatial DoF gains are ultimately overshadowed by escalating cumulative hardware losses and prohibitive power penalties.

\subsection{SE-Circuit Complexity, EE-Circuit Complexity, and EE-SE Tradeoffs}

To explicitly reveal the intrinsic tradeoffs between system performance and MiLAC architectures, we further present the SE-circuit complexity and EE-circuit complexity tradeoffs in Fig. \ref{fig:lossy32-SE-CCF-CCG} and Fig. \ref{fig:lossy32-EE-CCF-CCG}, respectively. In both figures, solid lines denote the transmitter-side MiLAC circuit complexity ($K_F$), while dashed lines represent the receiver-side MiLAC circuit complexity ($K_G$). As shown in Fig. \ref{fig:lossy32-SE-CCF-CCG}, the SE-oriented MiLACs exhibit a pronounced horizontal gap between $K_F$ and $K_G$. This massive gap visually corroborates our theoretical insight: pure SE maximization enforces a highly sparse transmitter-side MiLAC (low $K_F$) to prevent signal energy leakage, while tolerating a dense receiver-side MiLAC (high $K_G$) to exploit the noise attenuation effect for enhanced spatial DoF. In contrast, as shown in Fig. \ref{fig:lossy32-EE-CCF-CCG}, the EE-oriented MiLACs aggressively compress this horizontal gap, anchoring both $K_F$ and $K_G$ within an ultra-low regime even as $N_S$ increases. This confirms that the proposed EE optimization indiscriminately prunes power-hungry MiLAC interconnections at both the transmitter and receiver sides to strictly justify their power consumption.

Consequently, a compelling EE-SE tradeoff emerges between these two designs. As depicted in Fig. \ref{fig:lossy32-Tradeoff-SE-EE}, by tightly constraining $K_F$ and $K_G$ at higher $N_S$, the EE-oriented MiLACs sacrifice only a marginal fraction of SE (merely $2.5\%$ to $6.6\%$ for $N_S \in \{4,8,12,16\}$). However, this minor SE compromise translates into a substantial surge ($29.1\%$ to $84.7\%$) in EE across all $N_S$ settings, fundamentally bypassing the power consumption bottlenecks. Such significant improvements mainly arise from the reduced circuit complexity of both the transmitter-side and receiver-side MiLACs (by at most $26.4\%$ for the transmitter-side MiLAC and $59.8\%$ for the receiver-side MiLAC, respectively), as can be observed from \ref{fig:lossy32-SE-CCF-CCG} and Fig. \ref{fig:lossy32-EE-CCF-CCG}.

All in all, these results underscore the efficacy of the proposed LJAPOF: by enabling joint architecture and performance optimization, it dynamically synthesizes lossy MiLAC architectures that significantly outperform both stem-connected and fully-connected MiLACs in achieving optimal SE/EE, thereby pioneering the practical architecture design for lossy MiLACs and the deployment of lossy MiLAC-aided MIMO systems.

\section{Conclusion} \label{Sec. Con.}
In this paper, we investigated the joint architecture design for lossy MiLACs and SE/EE optimization in lossy MiLAC-aided MIMO systems. To combat the severe inter-stream interference and hardware losses inherently introduced by lossy TACs, we proposed a novel learning-based joint architecture and performance optimization framework, namely the LJAPOF. By integrating the differentiable MiLAC architecture generator, the physics-informed capacitance learning network, the parameter-free physics-constrained differentiable water-filling layer, and the dual-rate loss structure, the proposed LJAPOF successfully incorporates non-convex physical hardware constraints on lossy MiLACs into a stable end-to-end differentiable process, eliminating shortcut learning and the single-stream optimization trap, and ultimately enabling optimal architecture design and analog beamforming in lossy MiLAC-aided MIMO systems. Numerical results demonstrate that the proposed LJAPOF can maximize the SE/EE by intelligently balancing dense interconnections for interference suppression with sparse interconnections to mitigate power consumption and hardware losses. By navigating this tradeoff to consistently outperform stem-connected and fully-connected MiLACs, the proposed LJAPOF establishes a foundational paradigm for the practical architecture design of lossy MiLACs and deployment of future analog computing-aided MIMO networks.

\appendices

\section{Analysis on the Ohmic Power Dissipation of Lossy MiLACs}\label{App. A}

Let $\mathbf{Y}_F \in \mathbb{C}^{(N_S + N_T) \times (N_S + N_T)}$ and $\mathbf{Y}_G \in \mathbb{C}^{(N_R + N_S) \times (N_R + N_S)}$ denote the admittance matrices of the transmitter-side and receiver-side MiLACs, respectively. Taking the transmitter side as an example, the $N_S$ RF chains and $N_T$ antennas are modeled as the Norton equivalent current sources and load terminations, respectively. We define the termination admittance matrix as $\mathbf{Y}_{\text{term},F} = \text{diag}(Y_0 \mathbf{1}_{N_S}^T, Y_L \mathbf{1}_{N_T}^T)$, where $Y_0 = 1/Z_0$ and $Y_L = 1/Z_L$ are the internal conductance of the RF chains and the load conductance of the antennas, respectively. Typically $Z_0 = Z_L = 50\ \Omega$. The augmented system admittance matrix is given by $\tilde{\mathbf{Y}}_F = \mathbf{Y}_F + \mathbf{Y}_{\text{term},F}$. Further, let $P_{F,n}$, $n=1, \ldots, N_S$, denote the power allocated for the $n$-th input port. According to the Norton equivalent circuit, the injected root mean square (RMS) current vector is constructed as $\mathbf{i}_F = [i_{F,1}, \dots, i_{F,N_S}, 0, \dots, 0]^T \in \mathbb{C}^{N_S + N_T}$, where the magnitude of the $n$-th element is given by $\vert i_{F,n} \vert = \sqrt{4 P_{F,n} Y_0}$. Based on Kirchhoff's Current Law (KCL), the steady-state complex nodal voltage vector across the entire MiLAC network is obtained by solving $\mathbf{v}_F = \tilde{\mathbf{Y}}_F^{-1} \mathbf{i}_F$, which determines the exact Ohmic power dissipated by lossy TACs in the transmitter-side MiLAC as
\begin{align}
P_{\text{ohmic},F}(\mathbf{Y}_F) = \mathbf{v}_F^H \Re\{\mathbf{Y}_F\} \mathbf{v}_F, \notag
\end{align}
where $\Re\{\cdot\}$ extracts the real part of $\mathbf{Y}_F$ (i.e., the conductance matrix related to the power dissipation).

Similarly, the receiver-side MiLAC network operates with a physically inverted power flow, where the $N_R$ receive antennas act as the equivalent current sources capturing the incident electromagnetic waves, and the $N_S$ receive RF chains serve as the load terminations. Consequently, the termination admittance matrix for the receiver is defined as $\mathbf{Y}_{\text{term},G} = \text{diag}(Y_A \mathbf{1}_{N_R}^T, Y_0 \mathbf{1}_{N_S}^T)$, where $Y_A = 1/Z_A$ is the antenna radiation conductance (typically $Z_A = 50\ \Omega$). The augmented admittance matrix is $\tilde{\mathbf{Y}}_G = \mathbf{Y}_G + \mathbf{Y}_{\text{term},G}$. Let $P_{G,m}$ denote the received RF signal power at the $m$-th receive antenna. The equivalent Norton current excitation vector for the receiver-side MiLAC is formulated as $\mathbf{i}_G = [i_{G,1}, \dots, i_{G,N_R}, 0, \dots, 0]^T \in \mathbb{C}^{N_R + N_S}$, where the magnitude of the $m$-th element is given by $\vert i_{G,m} \vert = \sqrt{4 P_{G,m} G_A}$. Consequently, the nodal voltage vector across the receiver-side MiLAC network driven by the equivalent current excitation $\mathbf{i}_G$ is given by $\mathbf{v}_G = \tilde{\mathbf{Y}}_G^{-1} \mathbf{i}_G$ and the exact Ohmic power dissipated by the receiver-side lossy MiLAC, denoted by $P_{\text{ohmic},G}(\mathbf{Y}_G)$, can be modeled as
\begin{align}
P_{\text{ohmic},G}(\mathbf{Y}_G) = \mathbf{v}_G^H \Re\{\mathbf{Y}_G\} \mathbf{v}_G. \notag
\end{align}

\bibliographystyle{IEEEtran}
\bibliography{refer}

@book{pozar2011microwave,
  author    = {David M. Pozar},
  title     = {Microwave Engineering},
  edition   = {4},
  publisher = {John Wiley \& Sons},
  address   = {Hoboken, NJ, USA},
  year      = {2011},
  isbn      = {9780470631553}
}

@book{clerckx2013mimo,
  author    = {Clerckx, Bruno and Oestges, Claude},
  title     = {MIMO Wireless Networks: Channels, Techniques and Standards for Multi-Antenna, Multi-User and Multi-Cell Systems},
  publisher = {Academic Press},
  year      = {2013}
}

@inproceedings{he2016deep,
  title = {Deep Residual Learning for Image Recognition},
  booktitle = {Proc. {{IEEE}}/{{CVF}} CVPR},
  author = {He, Kaiming and Zhang, Xiangyu and Ren, Shaoqing and Sun, Jian},
  year = {2016},
  month = jun,
  pages = {770--778}
}

@article{geirhos2020shortcut,
  title = {Shortcut Learning in Deep Neural Networks},
  author = {Geirhos, Robert and Jacobsen, J{\"o}rn-Henrik and Michaelis, Claudio and Zemel, Richard and Brendel, Wieland and Bethge, Matthias and Wichmann, Felix A.},
  year = 2020,
  month = nov,
  journal = {Nat Mach Intell},
  volume = {2},
  number = {11},
  pages = {665--673},
  publisher = {Nature Publishing Group},
  copyright = {2020 Springer Nature Limited}
}

@article{nerini2025analog1,
  title = {Analog Computing for Signal Processing and Communications -- Part {{I}}: {{Computing}} with Microwave Networks},
  shorttitle = {Analog Computing for Signal Processing and Communications -- Part {{I}}},
  author = {Nerini, Matteo and Clerckx, Bruno},
  year = 2025,
  journal = {IEEE Trans. Signal Process.},
  volume = {73},
  pages = {5183--5197}
}

@article{nerini2025analog2,
  title = {Analog Computing for Signal Processing and Communications -- Part {{II}}: {{Toward}} Gigantic {{MIMO}} Beamforming},
  shorttitle = {Analog Computing for Signal Processing and Communications -- Part {{II}}},
  author = {Nerini, Matteo and Clerckx, Bruno},
  year = 2025,
  journal = {IEEE Trans. Signal Process.},
  volume = {73},
  pages = {5198--5212}
}

@article{peng2026lossy,
  title = {Lossy beyond Diagonal Reconfigurable Intelligent Surfaces: {{Modeling}} and Optimization},
  shorttitle = {Lossy beyond Diagonal Reconfigurable Intelligent Surfaces},
  author = {Peng, Yiyang and Li, Hongyu and Wu, Zheyu and Clerckx, Bruno},
  year = 2026,
  journal = {IEEE Trans. Wireless Commun.},
  volume = {25},
  pages = {7365--7380}
}

@article{nerini2025capacity,
  title = {Capacity of {MIMO} Systems Aided by Microwave Linear Analog Computers ({MiLACs})},
  author = {Nerini, Matteo and Clerckx, Bruno},
  journal = {arXiv preprint arXiv:2506.05983},
  year = {2025},
  month = jun
}

@article{nerini2026mimo,
  title = {{{MIMO}} Systems Aided by Microwave Linear Analog Computers: Capacity-Achieving Architectures with Reduced Circuit Complexity},
  shorttitle = {{{MIMO}} Systems Aided by Microwave Linear Analog Computers},
  author = {Nerini, Matteo and Clerckx, Bruno},
  year = 2026,
  journal = {IEEE Trans. Wireless Commun.},
  pages = {1--15}
}

@article{wu2026microwave,
  title = {Microwave Linear Analog Computer ({MiLAC})-Aided Multiuser {MISO}: Fundamental Limits and Beamforming Design},
  author = {Wu, Zheyu and Nerini, Matteo and Clerckx, Bruno},
  journal = {arXiv preprint arXiv:2601.10060},
  year = {2026},
  month = jan
}

@article{zhou2025beyonddiagonal,
  title = {Beyond-Diagonal {RIS} Under Non-Idealities: Learning-Based Architecture Discovery and Optimization},
  author = {Zhou, Binggui and Clerckx, Bruno},
  journal = {arXiv preprint arXiv:2510.15701},
  year = {2025},
  month = oct
}

@article{bengio2013estimating,
  title = {Estimating or Propagating Gradients through Stochastic Neurons for Conditional Computation},
  author = {Bengio, Yoshua and L{\'e}onard, Nicholas and Courville, Aaron},
  year = 2013,
  month = aug,
  journal = {arXiv preprint arXiv:1308.3432}
}

@article{ba2016layer,
  title={Layer normalization},
  author={Ba, Jimmy Lei and Kiros, Jamie Ryan and Hinton, Geoffrey E},
  journal={arXiv preprint arXiv:1607.06450},
  year={2016},
  month=jul
}

@inproceedings{choi2025deep,
  title = {Deep Learning-Based Power Allocation for Cell-Free Massive {{MIMO}} Networks with Adaptive Access Point Power Control},
  booktitle = {Proc. {{ICUFN}}},
  author = {Choi, Yoon-Ju and Yu, Ji-Hee and Jeong, Hye-Yoon and Kim, Ja-Eun and Song, Hyoung-Kyu},
  year = 2025,
  month = jul,
  pages = {744--749}
}

@inproceedings{perdana2021deep,
  title = {Deep Learning-Based Power Allocation in Massive {{MIMO}} Systems with {{SLNR}} and {{SINR}} Criterions},
  booktitle = {Proc. {{ICUFN}}},
  author = {Perdana, Ridho Hendra Yoga and Nguyen, Toan-Van and An, Beongku},
  year = 2021,
  month = aug,
  pages = {87--92}
}

@inproceedings{sanguinetti2018deep,
  title = {Deep Learning Power Allocation in Massive {{MIMO}}},
  booktitle = {Proc. ACSSC},
  author = {Sanguinetti, Luca and Zappone, Alessio and Debbah, Merouane},
  year = 2018,
  month = oct,
  pages = {1257--1261}
}

@article{wang2024reconfigurable,
  title = {Reconfigurable Intelligent Surface: {{Power}} Consumption Modeling and Practical Measurement Validation},
  shorttitle = {Reconfigurable Intelligent Surface},
  author = {Wang, Jinghe and Tang, Wankai and Liang, Jing Cheng and Zhang, Lei and Dai, Jun Yan and Li, Xiao and Jin, Shi and Cheng, Qiang and Cui, Tie Jun},
  year = 2024,
  month = sep,
  journal = {IEEE Trans. Commun.},
  volume = {72},
  number = {9},
  pages = {5720--5734}
}

@inproceedings{paszke2019pytorch,
  title = {{{PyTorch}}: An Imperative Style, High-Performance Deep Learning Library},
  shorttitle = {{{PyTorch}}},
  booktitle = {Proc. NeurIPS},
  author = {Paszke, Adam and Gross, Sam and Massa, Francisco and Lerer, Adam and Bradbury, James and Chanan, Gregory and Killeen, Trevor and Lin, Zeming and Gimelshein, Natalia and Antiga, Luca and Desmaison, Alban and Kopf, Andreas and Yang, Edward and DeVito, Zachary and Raison, Martin and Tejani, Alykhan and Chilamkurthy, Sasank and Steiner, Benoit and Fang, Lu and Bai, Junjie and Chintala, Soumith},
  year = 2019,
  volume = {32}
}

@article{li2026tutorial,
  title = {A Tutorial on Beyond-Diagonal Reconfigurable Intelligent Surfaces: Modeling, Architectures, System Design and Optimization, and Applications},
  shorttitle = {A Tutorial on Beyond-Diagonal Reconfigurable Intelligent Surfaces},
  author = {Li, Hongyu and Nerini, Matteo and Shen, Shanpu and Clerckx, Bruno},
  year = 2026,
  journal = {IEEE Commun. Surveys Tuts.},
  volume = {28},
  pages = {4086--4126}
}

@article{wu2025beyonddiagonal,
  title = {Beyond-Diagonal {{RIS}} in Multiuser {{MIMO}}: {{Graph}} Theoretic Modeling and Optimal Architectures with Low Complexity},
  shorttitle = {Beyond-Diagonal {{RIS}} in Multiuser {{MIMO}}},
  author = {Wu, Zheyu and Clerckx, Bruno},
  year = 2025,
  month = nov,
  journal = {IEEE Trans. Inf. Theory},
  volume = {71},
  number = {11},
  pages = {8506--8523}
}

@article{wu2026beyonddiagonal,
  title = {Beyond-Diagonal {{RIS}} Architecture Design and Optimization under Physics-Consistent Models},
  author = {Wu, Zheyu and Nerini, Matteo and Clerckx, Bruno},
  year = 2026,
  journal = {IEEE Trans. Wireless Commun.},
  volume = {25},
  pages = {14086--14100}
}

@article{shen2022modeling,
  title = {Modeling and Architecture Design of Reconfigurable Intelligent Surfaces Using Scattering Parameter Network Analysis},
  author = {Shen, Shanpu and Clerckx, Bruno and Murch, Ross},
  year = {2022},
  month = feb,
  journal = {IEEE Trans. Wireless Commun.},
  volume = {21},
  number = {2},
  pages = {1229--1243}
}

@article{li2024reconfigurable,
  title = {Reconfigurable Intelligent Surfaces 2.0: {{Beyond}} Diagonal Phase Shift Matrices},
  shorttitle = {Reconfigurable Intelligent Surfaces 2.0},
  author = {Li, Hongyu and Shen, Shanpu and Nerini, Matteo and Clerckx, Bruno},
  year = {2024},
  month = mar,
  journal = {IEEE Commun. Mag.},
  volume = {62},
  number = {3},
  pages = {102--108},
  copyright = {https://ieeexplore.ieee.org/Xplorehelp/downloads/license-information/IEEE.html}
}

@article{nerini2026physics,
  title={Physics-Compliant Modeling and Optimization of MIMO Systems Aided by Microwave Linear Analog Computers},
  author={Nerini, Matteo and Clerckx, Bruno},
  journal={arXiv preprint arXiv:2602.19379},
  year={2026}
}

@article{zhang2026channel,
  title={Channel Estimation in MIMO Systems Aided by Microwave Linear Analog Computers (MiLACs)},
  author={Zhang, Qiaosen and Nerini, Matteo and Clerckx, Bruno},
  journal={arXiv preprint arXiv:2601.11438},
  year={2026}
}

@article{abbas2017millimeter,
  title = {Millimeter Wave Receiver Efficiency: A Comprehensive Comparison of Beamforming Schemes with Low Resolution Adcs},
  shorttitle = {Millimeter Wave Receiver Efficiency},
  author = {Abbas, Waqas Bin and {Gomez-Cuba}, Felipe and Zorzi, Michele},
  year = 2017,
  month = dec,
  journal = {IEEE Trans. Wireless Commun.},
  volume = {16},
  number = {12},
  pages = {8131--8146}
}

@article{gast2024hardwareaware,
  title = {Hardware-Aware Energy Efficiency Optimization in Wireless Communications Using a Gearbox-{{PHY}}},
  author = {Gast, Florian and D{\"o}rpinghaus, Meik and Sen, Padmanava and Nimr, Ahmad and Fettweis, Gerhard P.},
  year = 2024,
  month = jul,
  journal = {IEEE Commun. Lett.},
  volume = {28},
  number = {7},
  pages = {1584--1588}
}

@inproceedings{mezghani2011modeling,
  title = {Modeling and Minimization of Transceiver Power Consumption in Wireless Networks},
  booktitle = {Proc. 2011 {{Int}}. {{ITG Workshop Smart Antennas}}},
  author = {Mezghani, Amine and Nossek, Josef A.},
  year = 2011,
  month = feb,
  pages = {1--8}
}

@article{zhang2019mixedadc,
  title = {Mixed-{{ADC}}/{{DAC}} Multipair Massive {{MIMO}} Relaying Systems: Performance Analysis and Power Optimization},
  shorttitle = {Mixed-{{ADC}}/{{DAC}} Multipair Massive {{MIMO}} Relaying Systems},
  author = {Zhang, Jiayi and Dai, Linglong and He, Ziyan and Ai, Bo and Dobre, Octavia A.},
  year = 2019,
  month = jan,
  journal = {IEEE Trans. Commun.},
  volume = {67},
  number = {1},
  pages = {141--153}
}

@inproceedings{chung200975gs,
  title = {A 7.5-{{GS}}/s 3.8-{{ENOB}} 52-{{mW}} Flash {{ADC}} with Clock Duty Cycle Control in 65nm {{CMOS}}},
  booktitle = {Proc. {{Symp}}. {{VLSI Circuits}}},
  author = {Chung, Hayun and Rylyakov, Alexander and Deniz, Zeynep Toprak and Bulzacchelli, John and Wei, Gu-Yeon and Friedman, Daniel},
  year = 2009,
  month = jun,
  pages = {268--269}
}

@inproceedings{lauridsen2016sleep,
  title = {Sleep Modes for Enhanced Battery Life of {{5G}} Mobile Terminals},
  booktitle = {Proc. {{IEEE}} {{VTC Spring}}},
  author = {Lauridsen, Mads and Berardinelli, Gilberto and Tavares, Fernando M. L. and Frederiksen, Frank and Mogensen, Preben},
  year = 2016,
  month = may,
  pages = {1--6}
}

@article{zhang2026quantization,
  title={Quantization-Aware {{EE}} Optimization and {{SE-EE}} Tradeoff for {{MiLAC}}-Aided {{MU-MISO}} Beamforming},
  author={Zhang, Yuchen and Zheng, Pinjun and Al-Naffouri, Tareq Y},
  journal={arXiv preprint arXiv:2604.24538},
  year={2026}
}

@article{peng2026joint,
  title={Joint Training Scattering Matrix Learning and Channel Estimation for Beyond-Diagonal Reconfigurable Intelligent Surfaces},
  author={Peng, Yiyang and Zhou, Binggui and Zheng, Yutong and Mandic, Danilo and Clerckx, Bruno},
  journal={arXiv preprint arXiv:2603.25299},
  year={2026}
}

@article{xiao2026deep,
  title = {Deep Learning-Based Channel Extrapolation for Dual-Band Massive {{MIMO}} Systems},
  author = {Xiao, Qikai and Li, Kehui and Zhou, Binggui and Ma, Shaodan},
  year = 2026,
  journal = {IEEE Wireless Commun. Lett.},
  volume = {15},
  pages = {2994--2998}
}

@article{zhou2024lowoverhead,
  title = {A Low-Overhead Incorporation-Extrapolation Based Few-Shot {{CSI}} Feedback Framework for Massive {{MIMO}} Systems},
  author = {Zhou, Binggui and Yang, Xi and Wang, Jintao and Ma, Shaodan and Gao, Feifei and Yang, Guanghua},
  year = 2024,
  month = oct,
  journal = {IEEE Trans. Wireless Commun.},
  volume = {23},
  number = {10},
  pages = {14743--14758}
}

@article{zhou2025beyond,
  title={Beyond-Diagonal RIS Under Non-Idealities: Learning-Based Architecture Discovery and Optimization},
  author={Zhou, Binggui and Clerckx, Bruno},
  journal={arXiv preprint arXiv:2510.15701},
  year={2025}
}

@article{sahinidis2019mixedinteger,
  title = {Mixed-Integer Nonlinear Programming 2018},
  author = {Sahinidis, Nikolaos V.},
  year = 2019,
  month = jun,
  journal = {Optim. Eng.},
  volume = {20},
  number = {2},
  pages = {301--306}
}

@article{caloz2013analog,
  title = {Analog Signal Processing: A Possible Alternative or Complement to Dominantly Digital Radio Schemes},
  shorttitle = {Analog Signal Processing},
  author = {Caloz, Christophe and Gupta, Shulabh and Zhang, Qingfeng and Nikfal, Babak},
  year = 2013,
  month = sep,
  journal = {IEEE Microw. Mag.},
  volume = {14},
  number = {6},
  pages = {87--103}
}
 
\end{document}